\documentclass[pre,showpacs,superscriptaddress,twocolumn,a4paper]{revtex4}
\usepackage[dvips]{graphicx}
\usepackage{epsfig,amsmath,amsfonts,amssymb,wasysym}

\begin{document}
\def\be{\begin{equation}}
\def\ee{\end{equation}}
\def\bfi{\begin{figure}}
\def\efi{\end{figure}}
\def\bea{\begin{eqnarray}}
\def\eea{\end{eqnarray}}

\newcommand{\reff}[1]{(\ref{#1})}
\def\op{{\varphi}}
\def\H{{\mathcal H}}
\def\rmd{{\rm d}}
\def\rme{{\rm e}}

\def\Cas{\mathrm{C}}
\def\bulk{\mathrm{bulk}}
\def\sur{\mathrm{surf}}
\def\ex{\mathrm{ex}}
\def\kB{k_{\mathrm{B}}}
\def\sf{\vartheta}
\def\Sf{\Theta}
\def\op{\phi}
\def\perm{\varepsilon}

\def\PP{{(+,+)}}
\def\PM{{(+,-)}}
\def\MM{{(-,-)}}
\def\MP{{(-,+)}}
\def\sp{{\rm (s,p)}}

\title{%
Critical Casimir effect in classical binary liquid mixtures%
}

\author{A.~Gambassi}
\altaffiliation{Present address: SISSA -- International School for Advanced
  Studies and INFN, via Beirut 2-4, 34151
  Trieste, Italy}
\affiliation{Max-Planck-Institut f\"ur Metallforschung,
Heisenbergstr.~3, D-70569 Stuttgart, Germany.} 
\affiliation{Institut f\"ur Theoretische und
Angewandte Physik, Universit\"at Stuttgart, Pfaffenwaldring~57, D-70569
Stuttgart, Germany.}
\author{A.~Macio\l ek}
\affiliation{Max-Planck-Institut f\"ur Metallforschung,
Heisenbergstr.~3, D-70569 Stuttgart, Germany.} 
\affiliation{Institut f\"ur Theoretische und
Angewandte Physik, Universit\"at Stuttgart, Pfaffenwaldring~57, D-70569 Stuttgart, Germany.}
\affiliation{Institute of Physical Chemistry, Polish Academy of
  Sciences, Kasprzaka 44/52, PL-01-224 Warsaw, Poland.}
\author{C.~Hertlein}
\affiliation{2.~Physikalisches Institut,
Universit\"at Stuttgart, Pfaffenwaldring~57, D-70569 Stuttgart, Germany.}
\author{U.~Nellen}
\affiliation{2.~Physikalisches Institut,
Universit\"at Stuttgart, Pfaffenwaldring~57, D-70569 Stuttgart, Germany.}
\author{L.~Helden}
\affiliation{2.~Physikalisches Institut,
Universit\"at Stuttgart, Pfaffenwaldring~57, D-70569 Stuttgart, Germany.}
\author{C.~Bechinger}
\affiliation{2.~Physikalisches Institut,
Universit\"at Stuttgart, Pfaffenwaldring~57, D-70569 Stuttgart, Germany.}
\affiliation{Max-Planck-Institut f\"ur Metallforschung,
Heisenbergstr.~3, D-70569 Stuttgart, Germany.} 
\author{S.~Dietrich}
\affiliation{Max-Planck-Institut f\"ur Metallforschung,
Heisenbergstr.~3, D-70569 Stuttgart, Germany.} 
\affiliation{Institut f\"ur Theoretische und
Angewandte Physik, Universit\"at Stuttgart, Pfaffenwaldring~57, D-70569
Stuttgart, Germany.}

\begin{abstract}
If a fluctuating medium is confined, the ensuing perturbation 
of its fluctuation spectrum
generates Casimir-like effective forces acting on its confining surfaces. Near
a continuous phase transition of such a medium the corresponding order
parameter fluctuations occur on all length scales and therefore close to the
critical point
this effect acquires a universal
character, i.e., to a large extent it is independent of the microscopic details
of the actual system. Accordingly it can be calculated theoretically by
studying suitable representative model systems. 
We report on the direct measurement of critical Casimir forces by total
internal reflection microscopy (TIRM), with 
femto-Newton resolution. 
The corresponding potentials are 
determined for individual colloidal particles floating above a substrate under
the action of the critical thermal noise in the solvent medium, 
constituted by a binary liquid mixture of water
and 2,6-lutidine near its lower consolute point.
Depending on the relative adsorption preferences of the colloid and substrate
surfaces with respect to the two components of the binary liquid mixture, we
observe that, upon approaching the critical point of the solvent, attractive
or repulsive forces emerge and
supersede those prevailing away from it.
Based on the knowledge of the critical Casimir forces acting in film
geometries within the Ising universality class and with equal or opposing
boundary conditions, we provide the corresponding theoretical predictions for
the sphere~---~planar 
wall geometry of the experiment. The experimental data for
the effective potential can be interpreted consistently in terms of these
predictions and a remarkable quantitative agreement is observed.
\end{abstract}


\maketitle

\section{Introduction} \label{sec:intro}

\subsection{Fluctuation-induced forces}

At macroscopic scales thermal or quantum fluctuations of a physical property 
of a system are typically negligible because 
fluctuations average out to zero upon increasing the length and time
scales at which the system is studied.  
At the micro- and nano-meter scale, instead, fluctuations become generally
relevant and, if externally controlled
and spatially confined, they give rise to novel phenomena. 
An example thereof is provided by the Casimir force acting on conducting
bodies~\cite{Casimir}, which is due to the confinement of 
\emph{quantum} fluctuations of the
electromagnetic field in vacuum 
and which influences the behavior of micrometer-sized
systems ranging from colloids to micro- and nano-electromechanical systems
(MEMS, NEMS). 

\emph{Thermal} 
fluctuations in condensed matter typically occur on a molecular
scale. However, upon approaching the critical point 
of  a second-order phase transition 
the fluctuations of the order parameter $\op$ 
of the phase transition become relevant and detectable at a much larger length
scale $\xi$ and their confinement results in a fluctuation-induced Casimir
force $f_\Cas$ acting on the confining surfaces~\cite{FdG}. 
This so-called critical Casimir force $f_\Cas$ has a range which is 
set by the correlation length $\xi$ of the fluctuations
of the order parameter. Since near the critical point 
$\xi$ can reach up to macroscopic
values, the range of $f_\Cas$ can be controlled and varied to a large extent 
by minute temperature changes close to the critical temperature $T_c$. 
We shall show that this control
of the thermodynamic state of the system is a manageable task. This implies
that the critical Casimir force can be easily switched on and off, which
allows one to identify it relative to the omnipresent background forces. 
In addition, by proper \emph{surface} treatments of 
the confining surfaces, the force can be relatively easily
turned from attractive to repulsive~\cite{krech:99:0,Dbook}
in contrast to the
Casimir force stemming from electromagnetic fluctuations for which such a
change requires carefully chosen \emph{bulk} materials providing 
the solid walls and the
fluid in between~\cite{Cap:2009}. 
Such a repulsive force might be exploited to prevent stiction in MEMS and
NEMS, which would open significant perspectives for applications. 
Finally, at $T_c$ the strength of the critical Casimir force can easily
compete with or even dominate dispersion forces, with which it shows the same
algebraic decay, however without suffering from the weakening due to
retardation effects. 
The universality of $f_\Cas$ means that the same force is
generated near the critical point of liquid-vapor coexistence of \emph{any}
fluid or near the consolute point of phase segregation of \emph{any} binary
or multicomponent liquid mixture. This allows one to pick and use those
representatives of the universality class which in addition optimize desired
performances of MEMS and NEMS. This provides a highly welcome flexibility. 

The fluctuation-induced forces generated by confining the
fluctuations of electromagnetic fields in the quantum vacuum
(Casimir effect) or of the order parameter in a critical medium (critical
Casimir effect) have a common description within the field-theoretical
approach. Accordingly, the connection  between these two effects
goes well beyond the mere analogy and it indeed becomes an exact mapping
in some specific cases of spatial dimension $d$, geometries, 
and boundary conditions. 
This deep connection actually justifies the use of the term ``critical
Casimir force'' when referring to the effective force due
to the confinement of critical fluctuations. 
On the other hand, from a theoretical point of view the quantum and the
critical Casimir effect are also distinct in that the quantum one in vacuum
corresponds to a free field theory whereas the critical one is described by a
more challenging non-Gaussian field theory.

\subsection{Finite-size scaling}

The theory of finite-size scaling (see, e.g., Refs.~\cite{krech:99:0,Dbook}) 
predicts that in the vicinity of $T_c$ the  {\it critical} Casimir force
$f_\Cas$ and its dependence on
temperature are described by a {\it universal} scaling function 
which depends only on the gross features of the system and of the confining
surfaces, i.e., on the so-called universality class of the phase transition
occurring in the bulk and on the geometry and surface universality
classes of the confining surfaces~\cite{Binder83,diehl:86:0,diehl:97}.
The latter characterize the boundary conditions 
(BC)~\cite{Binder83,diehl:86:0,diehl:97,krech:99:0} 
the surfaces impose on the fluctuations 
of the order parameter of the underlying second-order phase transition.
The actual physical nature of the order parameter $\op$ depends on which kind
of continuous 
phase transition is approached: in the case we shall be mainly
concerned with in the following, i.e., the consolute point of phase
segregation in binary liquid mixtures, $\op$ is
given by the difference between the local and the mean concentration of one
of the two components of the mixture (see, c.f., 
Sec.~\ref{sec:exp_bm} for further details). 
For binary liquid mixtures the confining surfaces generically exhibit
preferential adsorption of one of the two components of the mixture,
resulting in an enhancement of the order parameter $\phi$ close to the surface.
(This amounts to the presence of symmetry-breaking surface fields, see, e.g., 
Refs.~\cite{Binder83,diehl:86:0,diehl:97}.) 
One usually refers to the corresponding boundary conditions as $(+)$ or $(-)$
depending on whether the surface favors $\op > 0$ or $\op <0$, respectively.
Due to its universal nature, the critical Casimir force  can be studied via
representative models which are amenable to theoretical investigations. 
Since due to universality microscopic details can only in a rather limited
way be blamed for potential discrepancies, the resulting predictions face very
stringent experimental tests. 

Most of the available theoretical and experimental studies 
focus on the {\it film geometry} in which the system undergoing the
second-order phase transition is confined between two parallel surfaces of
large transverse area $S$ at a distance $L$.  
For this geometry and assuming that the only relevant thermodynamic
variable is the temperature $T$ (possible additional variables such as the
concentration are set to their
critical values), renormalization-group theory 
shows~\cite{krech:92a,krech:92b} 
that the critical Casimir force $f_\Cas$ scales as
\be 
\label{eq:scf}
\frac{f_\Cas(T,L)}{\kB T} = 
\frac{S}{L^3} \sf(\tau(L/\xi^+_0)^{1/\nu}) 
\ee
in three spatial dimensions ($d=3$), 
where $\sf(x)$ is a universal scaling function, and 
$\tau$ is the reduced deviation from the critical
temperature $T_c$ such that $\tau > 0$ corresponds to the disordered
(homogeneous) phase. If, as it is usually the case, the homogeneous phase
is located at high temperatures in the phase diagram of the system, one
defines $\tau = (T-T_c)/T_c$. 
However, there are also cases -- such as the one we shall be interested in
(see, c.f., Fig.~\ref{fig:WL-PD}) -- in which this phase is located at low
temperatures so that there one defines $\tau = -(T-T_c)/T_c$.    
The system-specific (i.e., non-universal)  amplitudes $\xi_0^\pm$ in
Eq.~\reff{eq:scf} enter into the algebraic behavior of the {\it bulk}
correlation length $\xi$  of the order parameter $\op$ upon approaching the
critical point:
\be 
\xi(\tau \rightarrow 0^\pm)=\xi^\pm_0 |\tau|^{-\nu} \;.
\label{eq:xidiv}
\ee
In what follows we shall mainly consider $\xi_0 \equiv \xi_0^+$, which
forms with $\xi_0^-$ a {\it universal} amplitude ratio $U_{\xi_{\rm gap}}
\equiv \xi_0^+/\xi_0^-\simeq 1.9$~\cite{PV,Priv} 
in those cases in which $\xi(\tau < 0)$ is finite.
(Renormalization-group theory tells that in the bulk there are only two
independent non-universal amplitudes, say $\xi_0^+$ and 
$C_\phi = \langle\phi\rangle/(-\tau)^\beta$
of the order parameter below $T_c$; all other non-universal amplitudes can be
expressed in terms of them and universal amplitude ratios~\cite{Priv}. Here
$\beta$ is the critical exponent which characterizes the singular behavior
$\langle\phi\rangle\sim(-\tau)^\beta$  of the average order parameter
$\langle\phi\rangle$  for $\tau\rightarrow 0^-$, with $\beta = 0.3265(3)$ for
the three-dimensional Ising universality class~\cite{PV}.)
The bulk correlation length $\xi$ can be inferred from, e.g., the 
exponential decay of the two-point correlation function 
of the order parameter. The algebraic increase of $\xi$ [Eq.~\reff{eq:xidiv}] 
is characterized by the universal
exponent $\nu$ which equals $0.6301(4)$ for the three-dimensional Ising
universality class~\cite{PV} which captures, among others, the
critical behavior of binary liquid mixtures close to the demixing
point as studied experimentally here.

\subsection{Theoretical predictions and previous experiments}

For the Ising universality class with symmetry-breaking boundary conditions
theoretical predictions for the universal scaling function $\sf$ are available
from field-theoretical~\cite{krech,upton} 
and Monte Carlo studies~\cite{krech,vas-07,vas-08}.
The critical Casimir force turns out to be {\it attractive} for equal
boundary conditions (BC) on the two surfaces, i.e., $\PP$ or $\MM$, 
whereas it is
{\it repulsive} and generically stronger for opposing boundary conditions,
i.e., $\PM$ or $\MP$.
In the presence of such boundary conditions, for 
topographically~\cite{patt-top} 
or chemically~\cite{patt-chem}
patterned confining surfaces or for curved 
surfaces~\cite{colloids1a,colloids1b}
theoretical 
results are available primarily within mean-field theory.

Following theoretical predictions and suggestions~\cite{krech:92c}, 
previous {\it indirect} evidences for both attractive and repulsive critical
Casimir force were based on studying fluids close to critical
endpoints (see Ref.~\cite{gam-08} for a more detailed summary). 
Under such circumstances, 
the film geometry with parallel planar walls can be indeed 
experimentally realized by forming complete
wetting fluid films~\cite{diet:98} 
in which a liquid phase is confined between a solid
substrate (or another spectator phase) 
and the interface with the vapor phase and its thickness $L$
can be tuned by undersaturation, in particular off criticality. 
Upon changing 
pressure and temperature one can drive the
liquid film towards a second-order phase transition which 
nonetheless keeps the
confining liquid-vapor interface sharp. 
The fluctuations of the associated order
parameter, confined within the film of thickness $L$, give rise to
a critical Casimir pressure (related to $\sf$~\cite{krech:92c}) 
which acts on the liquid-vapor interface,
displacing it from the equilibrium position it would have under the
effect of dispersion forces alone, i.e., in the absence of critical
fluctuations. This results in a temperature-dependent change of $L$. 
Based on the knowledge of the relationship between $L$ and pressure, 
by monitoring this variation it is possible to infer indirectly the
magnitude of the Casimir force which drives this change of thickness.
This approach has been used for the study of wetting films of $^4$He at
the normal-superfluid transition~\cite{garcia4}, 
for $^3$He-$^4$He mixtures close to the
tricritical point~\cite{garcia3},  
and for classical binary liquid mixtures close to
demixing transitions~\cite{pershan,rafai}. 
The film thickness $L$ has been
determined by using capacitance~\cite{garcia4,garcia3} or 
X-ray reflectivity measurements~\cite{pershan}, or ellipsometry~\cite{rafai}.
For the results of Refs.~\cite{garcia4}, \cite{garcia3}, and~\cite{pershan}
the quantitative agreement with the theoretical 
predictions for the corresponding bulk and surface universality 
classes 
(see
Refs.~\cite{krech:92a,krech:92b,krech:92c,vas-07,vas-08,hucht,hasen,DK,kardar:04,LGW-MFa,LGW-MFb},
\cite{LGW-MFa,MD:06}, and~\cite{upton,vas-07,vas-08,LGW-MFa}, respectively)
are excellent~\cite{garcia4} 
or remarkably good~\cite{garcia3,pershan}. 
For $^4$He~\cite{garcia4} one has Dirichlet-Dirichlet boundary conditions, for
$^3$He-$^4$He mixtures~\cite{garcia3} Dirichlet-$(+)$ boundary conditions, and
in Ref.~\cite{pershan} $\PM$ boundary conditions hold. 

\subsection{Direct determination of critical Casimir forces}

The aim of the experimental investigation discussed here 
is to provide a {\it direct} determination of the Casimir force, by measuring
the associated potential $\Phi_C$.
On dimensional grounds and on the basis of Eq.~\reff{eq:scf}, the
scale of this potential is set by $\kB T_c$ and therefore, as realized in
Ref.~\cite{pershan}, in order to enhance the strength of the critical Casimir
force it is desirable to engage critical points with higher
$T_c$ compared to those of the $\lambda$-transition investigated in
Refs.~\cite{garcia3,garcia4}. This consideration suggests {\it classical} 
fluids as natural candidates for the critical medium. The experimentally
driven preference for having $T_c$ and the critical pressure to be close to
ambient conditions can be satisfied by numerous binary liquid mixtures which
exhibit consolute points for phase segregation.
From Eq.~\reff{eq:scf} one can infer a rough estimate of the critical 
Casimir
force $f_\Cas$. For an object which exposes an effective area $S = 1\,\mu\mbox{m}^2$ to a wall at a distance $L=100\,\mbox{nm}$, and for
$T_c=300\,\mbox{K}$ one finds $f_\Cas \lesssim 4\,\mbox{pN}$. Since the
scaling function $\sf(x)$ vanishes upon moving away from criticality, i.e., 
$\sf(|x|\rightarrow\infty) \rightarrow 0$, and because one is interested in
probing also larger distances $L$, one needs force measurements with a force
resolution which is significantly better than pN. Atomic force microscopy at
room temperature cannot deliver fN accuracy. 
This required sensitivity can, however, 
be achieved by using total internal reflection microscopy (TIRM),
which enables one to determine the potential of the effective forces 
acting on a colloidal particle near a wall, 
by monitoring its Brownian motion in a solvent.   
Choosing as the solvent a suitable 
binary liquid mixture allows one to investigate the critical Casimir force on
the particle which arises upon approaching the demixing transition of the
mixture. 
Such a second-order phase transition falls into the bulk
universality class of the Ising model.
In this geometrical setting the fluctuation spectrum of the critical 
medium (i.e., the binary liquid mixture) 
is perturbed by the confinement due to a flat wall and by the presence of the
spherical cavity. The curvature of one of the two confining surfaces
introduces an additional length scale and thus leads to an extension of 
the scaling form in Eq.~\reff{eq:scf} such
that the scaling function $\sf$  additionally depends on the ratio between the
radius $R$ of the colloid and the minimal distance $z$ between the surface of
the colloid and the flat surface of the substrate [c.f.,
Sec.~\ref{sec:th:cc}; here $z$ plays the role of $L$ in 
Eq.~\reff{eq:scf}].
At present, for arbitrary values of $z$ and radii of curvature, 
theoretical predictions for the critical Casimir force in a
geometrical setting involving one non-planar surface are available only within
mean-field theory, both for spherical~\cite{colloids1a,colloids1b} 
and ellipsoidal~\cite{khd-08}
particles, which demonstrate that the results of the so-called 
Derjaguin approximation are valid for $z/R \ll
1$~\cite{colloids1a,colloids1b} (see, c.f., Sec.~\ref{sec:th}).
Beyond mean-field theory and for various universality classes, theoretical
results have been 
obtained in the so-called protein limit corresponding to $z/R$,
$\xi/z \gg 1$~\cite{eisen,krech:92a}
where $R$ indicates the typical size of the, in general  
nonspherical, particle.
However, at present this protein limit is not accessible by TIRM because for
small particles far away from the substrate 
(through which the evanescent optical
field enters into the sample) the signal of the scattered light from the
particle is too weak. The experimentally relevant case is the opposite one of
a large colloidal particle close to the wall. Although in $d=3$ theoretical
results for the full scaling function of the sphere-plate geometry are not
available, in this latter case one can 
take advantage of the Derjaguin approximation in order to express the
critical Casimir force $F_\Cas$ acting on the colloid in terms of the
force acting within a film geometry, which was investigated successfully 
via Monte Carlo simulations in Refs.~\cite{vas-07,vas-08}. 
This is explained in detail in
Sec.~\ref{sec:th:cc}, 
in which we present the theoretical predictions for the
scaling function of the critical Casimir force (and of the associated
potential) for the case of a sphere near a wall immersed into a binary liquid 
mixture at its critical composition. On the other hand, 
in Sec.~\ref{sec:th:nc} we discuss
the expected behavior of the effective potential of the colloid if the
binary liquid mixture is not at its critical concentration so that, upon
changing the temperature, it undergoes a first-order phase transition. 
The discussions in
Sec.~\ref{sec:th} form the basis for the interpretation of the experimental
results. The experimental setting is described in Sec.~\ref{sec:exp}. In
Sec.~\ref{sec:exp_TIRM}
we recall the principles of TIRM and of the data analysis, 
whereas in Sec.~\ref{sec:exp_bm} we discuss
the specific choice of the binary mixture used here and how one can 
experimentally realize the various boundary conditions. In
Sec.~\ref{sec:res} we present in detail the experimental results, comparing
them with the theoretical predictions, for mixtures both at critical and
non-critical compositions. A summary and a discussion of perspectives and of
possible applications of our findings are provided in Sec.~\ref{sec:conc}.
Part of the analysis presented here has been reported briefly in
Ref.~\cite{nature}. 
(For a pedagogical introduction to the subject see Ref.~\cite{EPN}.)

\section{Theoretical predictions} 
\label{sec:th}

\subsection{Critical composition}
\label{sec:th:cc}

\subsubsection{General properties}
The critical Casimir force $F_\Cas$ acting on a spherical particle of radius
$R$, at a distance $z$ of closest approach from the flat surface of a
substrate and  immersed in a near-critical medium at temperature $T\simeq T_c$
takes, for strong preferential adsorption, 
the \emph{universal} scaling form~\cite{colloids1a,colloids1b,colloids2}
\be
F_\Cas(z) = \frac{\kB
  T}{R}K_\pm^{\rm (s,p)}\left(x\equiv\frac{z}{\xi},\Delta\equiv\frac{z}{R}\right) \,.
\label{eq:Ksp}
\ee
The scaling function $K_\pm^{\rm (s,p)}(x,\Delta)$ depends, in addition, 
on the combination of (sphere, plate) [(s,p)] boundary conditions imposed by
the surfaces of the sphere and of the plate and on the phase from which the
critical point is approached (i.e., on the sign of $\tau$, with 
$K_\pm^{\rm (s,p)}$ corresponding to $\tau \gtrless 0$). 
(In line with Eq.~\reff{eq:xidiv} and with the standard notation in the
literature, the one-phase region is denoted by $+$ and the two-phase region by
$-$. These signs should not be confused with the signs $\PP$ etc. indicating,
also in line with the literature, the character of the boundary conditions of
the two confining surfaces (s,p). In order to avoid a clumsy notation we
suppress or use these two notations in a selfevident way.)
The scaling
form of the associated potential  
$\Phi_C(z) \equiv \int_z^\infty\!\!\rmd s\; F_\Cas(s)$ 
follows by integration of Eq.~\reff{eq:Ksp}. In the two limiting cases $\Delta
\gg 1$ and $\Delta \ll 1$ it is possible to 
calculate $K(x,\Delta)$ on the basis of the
so-called small-sphere expansion and Derjaguin approximation,
respectively~\cite{colloids1a,colloids1b,colloids2}. 
In the former case one finds in three space dimensions, for $\tau>0$ and
symmetry-breaking boundary conditions $\sp = (\pm,+)$ (see
  Eq.~(7) in Ref.~\cite{colloids1a}, which also includes higher-order terms)
\be
\begin{split}
K_+^{(\pm,+)}(x,\Delta \rightarrow \infty) = & \mp \frac{a}{c_+} \frac{x^{\beta/\nu+1}}{2^{\beta/\nu}} P_+'(x)
  \Delta^{-(\beta/\nu+1)} \\
& + O(\Delta^{-2\beta/\nu-1})\,,
\end{split}
\label{eq:Ksfe}
\ee
where $\beta/\nu \simeq 0.518$. In this limit, the force acting on the
``small'' particle is determined, to leading order, by the interaction
between the particle and  
the average order parameter profile $\langle\phi(z)\rangle_{\infty/2}$ 
induced by the planar
wall in the absence of the particle, i.e., in a semi-infinite system 
($\infty/2$). This profile 
is characterized for $\tau>0$ by the \emph{universal} 
scaling function $P_+$ entering
Eq.~\reff{eq:Ksfe}: $\langle\phi(z)\rangle_{\infty/2} = 
\langle\phi\rangle_{\infty,-\tau<0} P_+(z/\xi)$, where
$\langle\phi\rangle_{\infty,-\tau<0} = C_\phi \tau^\beta$ 
is the value of the order parameter in the bulk ($\infty$) 
corresponding to the reduced temperature
$-\tau\rightarrow 0^-$. The universal constant $c_+$ in
Eq.~\reff{eq:Ksfe} characterizes the critical adsorption profile
$P_+(x\rightarrow 0) \rightarrow c_+ x^{-\beta/\nu}$, whereas
$a=A_\phi^2/B_\phi$ is the \emph{universal} ratio~\cite{colloids2} between the
non-universal amplitudes $A_\phi$ and $B_\phi$ of the critical order parameter
profile in the semi-infinite system $\langle\phi(z)\rangle_{\infty/2, \tau=0}
= A_\phi (2 z)^{-\beta/\nu}$ and of 
the two-point correlation function in the bulk
$\langle\phi({\bf r})\phi(0)\rangle_{\infty,\tau=0} = B_\phi r^{-2\beta/\nu}$,
respectively. In turn, $A_\phi$ (and therefore $B_\phi$) can be expressed in
terms of the two independent non-universal amplitudes $\xi_0^+$ and $C_\phi$
via $A_\phi = c_+ C_\phi/(2\xi_0^+)^{-\beta/\nu}$. (For a detailed discussion
of the values of these universal amplitude ratios we refer the reader to
Refs.~\cite{colloids1a,colloids1b,colloids2}.)

\subsubsection{Derjaguin approximation}
Equation~\reff{eq:Ksfe} is useful to discuss the behavior of
colloids which are small compared to their distance from the plate.
However, in the experiment discussed in Sec.~\ref{sec:exp}, the distance $z$
is typically much smaller than the radius $R$ of the particle. This case can
be conveniently discussed within the Derjaguin approximation, which yields in
three dimensions~\cite{colloids1a}
\be
K(x,\Delta \rightarrow 0) = \Delta^{-2}\hat\sf(x)\,,
\label{eq:KDerj}
\ee
where the expression for $\hat\sf(x)$ is determined further below in terms of
the scaling function $\sf$ of the critical Casimir force $f_\Cas$ acting
within a film [see Eq.~\reff{eq:scf}].

\begin{figure}
\centering\includegraphics[width=\columnwidth]{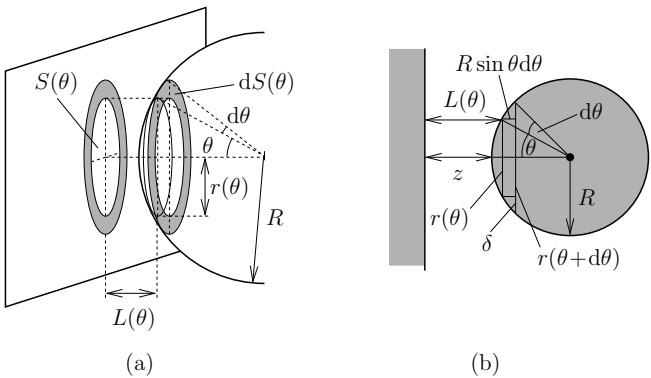}
\caption{%
(a) Geometry of the Derjaguin approximation for the plate-sphere geometry, (b)
  cross section through the center of the sphere and normal to the plate. In
  (a), the base of a cap of the sphere with radius $r(\theta)$ is shown as a
  thin line. The inner circle of the grey ring $\rmd S(\theta)$ is this thin
  line shifted by $R\sin\theta\rmd\theta$ towards the center of the sphere. In
  (b), $\delta = r(\theta+\rmd\theta) - r(\theta)$ is the cross section of the
  grey ring $\rmd S(\theta)$ shown in (a).%
}
\label{fig:Derj}
\end{figure}

The scaling functions $\sf_{\PM}(x)$ and $\sf_{\PP}(x)$ 
for the boundary conditions $\PM$ and $\PP$ relevant to the
study of the critical properties of binary liquid mixtures at their critical
compositions have been determined by Monte Carlo
simulations~\cite{vas-07,vas-08}. 
Within the  Derjaguin approximation, valid 
for $\Delta \ll 1$, i.e., if the radius $R$ of the colloid is much larger than
the minimal separation $z$ between
the surface of the colloid and the flat substrate, 
the curved surface of the colloid
is considered to be made up of successive circular rings of infinitesimal 
area $\rmd S(\theta)$ and radius $r(\theta)$ which are parallel
to the substrate and are at a normal distance 
$L(\theta) = z + R (1-\cos\theta)$ 
from an opposing identical circular ring on the surface of the substrate (see 
Fig.~\ref{fig:Derj}).
%
%
%
%
Assuming additivity, the contribution $\rmd F_\Cas$ 
of each single pair of rings
to the total Casimir force $F_\Cas$ is given by
\be
\frac{\rmd F_\Cas}{\kB T} = \frac{\rmd S(\theta)}{L^3(\theta)} \sf(L(\theta)/\xi)
\ee
where $\sf$ is the scaling function of the critical Casimir force acting
within the \emph{film} 
geometry [see Eq.~\reff{eq:scf}]. Here it is convenient to
express $\sf$ not as a function of $u \equiv \tau (L/\xi_0)^{1/\nu}$ as in
Eq.~\reff{eq:scf} but as a function of $x = L/\xi$ where $x = u^{\nu}$ for
$u>0$ and $x= U_{\xi_{\rm gap}}(-u)^{\nu}$ for $u<0$  with 
$U_{\xi_{\rm gap}}\simeq
1.9$~\cite{PV} for the three-dimensional Ising universality class we are
interested in.
The radius $r(\theta)$ of the ring is given by 
$r(\theta) = R \sin\theta$ and therefore its area is $\rmd S(\theta)
= \pi [r(\theta+\rmd\theta)]^2 - \pi [r(\theta)]^2 = 
2 \pi R^2 \sin\theta\cos\theta\,\rmd\theta$. 
The total force $F_\Cas$ is obtained by summing all the contributions $\rmd
F_\Cas(\theta)$ of the circular rings, up to the maximal angle
$\theta_{\rm M}$
\be
\frac{F_\Cas}{\kB T} = \int_{\theta = 0}^{\theta_{\rm M}}\; \frac{\rmd
  S(\theta)}{L^3(\theta)} \sf(L(\theta)/\xi)\,. 
\ee 
$\theta_{\rm M}=\pi/2$ is a natural choice, neglecting any influences from the
back side of the sphere. However, we shall see below that its specific 
value does not affect the result in the limit $R \gg z$.
For $R \gg z$ the integral (due to the denominator) is dominated by the
contributions it picks up at 
small angle $\theta$, so that one can approximate $L(\theta)/z
\simeq 1 + (R/z)(\theta^2/2)$ and therefore
\be
\frac{F_\Cas(z)}{\kB T} = \frac{2\pi R^2}{z^3} \int_0^{\theta_{\rm M}}\rmd\theta\;
\frac{\theta}{\left[1 + \frac{R}{z}\theta^2/2\right]^3}  
\sf([1 + \frac{R}{z}\theta^2/2] z/\xi) \, .
\ee
[For $\theta_{\rm M}\rightarrow \infty$ this is identical with Eq.~(4) in
  Ref.~\cite{colloids1a}.]
Introducing the variable $l = 1 +(R/z)(\theta^2/2)$, 
one can write the previous expression as
\be
\frac{F_\Cas(z)}{\kB T} = \frac{2\pi R}{z^2} \int_1^{l_{\rm
    M}}\rmd l\; \frac{1}{l^3} \sf(l\,  z/\xi)
\ee
where $l_{\rm M} \equiv 1 + (R/z) (\theta_{\rm M}^2/2)$. 
In the limit $R/z \rightarrow \infty$, $l_{\rm M} \rightarrow \infty$ 
independently of $\theta_{\rm M}$, so that the
integral can be extended up to $\infty$ and
\be
\frac{F_\Cas(z)}{\kB T} = \frac{R}{z^2} \hat\sf(z/\xi)
\label{eq:scalF}
\ee
where
\be
\hat\sf(x) =  2\pi  \int_1^\infty \rmd l\; \frac{1}{l^3} \sf(l\, x).
\label{eq:scalingfuncF}
\ee 
The potential $\Phi_C(z)$ associated with
the Casimir force is given by
\be
\begin{split}
\frac{\Phi_C(z)}{\kB T} &
= \frac{2 \pi R}{z} \int_1^\infty\!\!\rmd y
\int_1^\infty \!\!
\rmd l\; y^{-2} l^{-3} \sf(l y \, z/\xi) \\
&= \frac{R}{z} \; 2\pi\int_1^\infty\rmd v
\left(\frac{1}{v^2} - \frac{1}{v^3} \right) \sf (v\, z/\xi)\\
&=  \; \frac{R}{z} \; \Theta(z/\xi)
\end{split}
\label{eq:PhiDgen}
\ee 
where we have changed the 
variable $l\mapsto v \equiv l y$, exchanged the order
of the remaining integrals $\int_1^\infty\rmd y\int_y^\infty\rmd v =
\int_1^\infty\rmd v\int_1^v\rmd y$, and introduced the scaling function
\be
\Theta(x) \equiv  2\pi\int_1^\infty\rmd v
\left(\frac{1}{v^2} - \frac{1}{v^3} \right) \sf (v\, x) \, .
\label{eq:scalingfunc}
\ee
According to Eqs.~\reff{eq:scalF} and~\reff{eq:PhiDgen}, 
for separations $z$ much smaller than the radius of the colloid, the Casimir
force and the Casimir potential increase linearly upon 
increasing the radius $R$ of the colloid. 
At the bulk critical point, $\hat\sf(0) = \Theta(0) = \pi \sf(0)$
and $\hat\sf'(0) = 2 \pi \sf'(0)$, whereas
$\Theta'(0) = \infty$. If in the film geometry the force is attractive
(repulsive) at all temperatures, within the Derjaguin approximation the same
sign holds also in the sphere-plate geometry.
At the critical concentration 
the Casimir force acting on a $(+)$ colloid in front of a $(-)$ substrate is
the same as the one acting on a  $(-)$ colloid in front of a $(+)$
substrate. This is no longer true for non-critical concentrations. 
Although the Derjaguin approximation is expected to be valid only 
for $R \gg z$, the comparison between the results of the mean-field
calculation~\cite{colloids1a,colloids1b} for the actual sphere-plate geometry
and the ones of the corresponding 
Derjaguin approximation
based on the mean-field theory (MFT) 
for the film geometry show good agreement even for
$z/R$ up to $0.4\div0.5$. 

\subsubsection{Theoretical predictions for scaling functions}
\begin{figure}
\centering\includegraphics[width=\columnwidth]{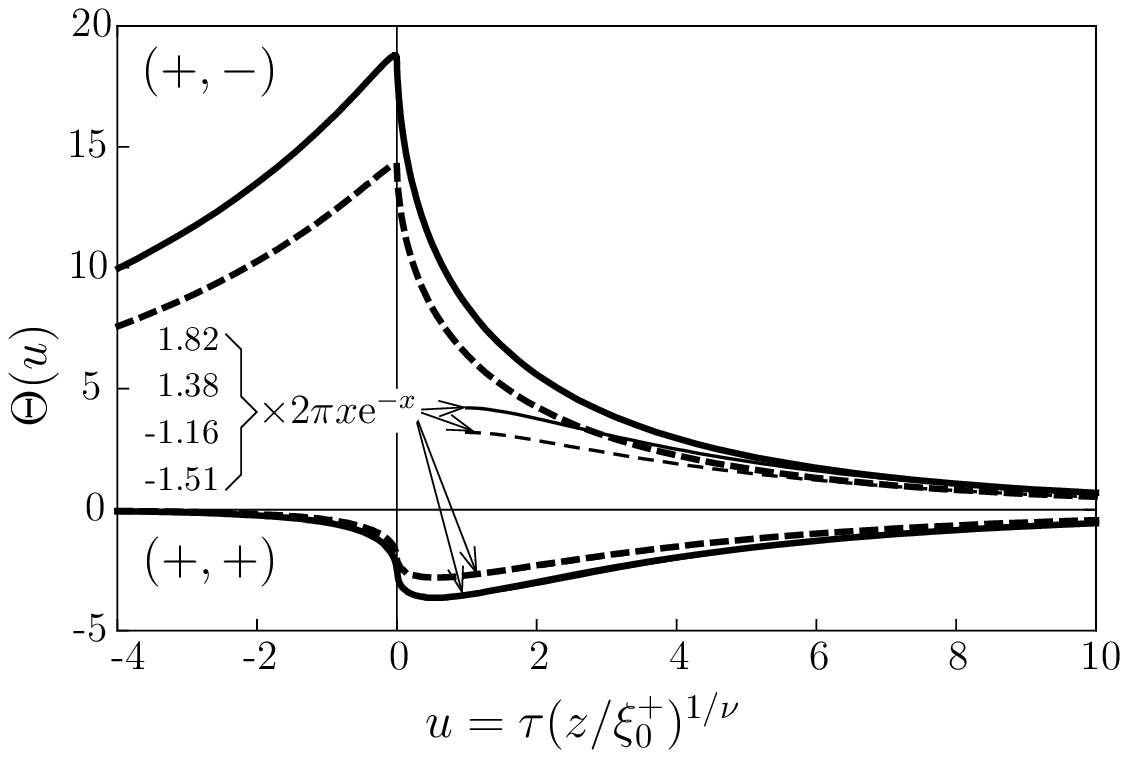}
\caption{Scaling functions  $\Sf_\PP$  and $\Sf_\PM$ of the Casimir potential
  $\Phi_C$ [see Eq.~\reff{eq:PhiDgen}] 
  for $\PP$ and $\PM$ BC, respectively, within the Derjaguin
  approximation and for the three-dimensional Ising universality class, as
  functions of $u = \tau (z/\xi_0^+)^{1/\nu}$ with $\nu \simeq 0.630$. 
The thick 
solid and dashed lines have been obtained via Eq.~\reff{eq:scalingfunc} 
on the basis of the Monte Carlo
estimates for $\sf_{\PP,\PM}$ presented in 
Refs.~\protect{\cite{vas-07,vas-08}}, indicated by $(i)$ and $(ii)$,
  respectively, in Figs.~9 and 10 of Ref.~\cite{vas-08}.  
(The thick solid lines agree with the estimates reported in Fig.~2(d) of
  Ref.~\protect{\cite{nature}}.)
$\Theta_{\PP}$ attains its minimum value $\Theta_{\PP}^{\rm (min)}
\simeq -3.6$ (solid line) and $-2.8$ (dashed line) both for 
$u_{\rm min} \simeq 0.54$, whereas $\Theta_{\PM}$ attains (smoothly)
its maximum value $\Theta_{\PM}^{\rm (max)} \simeq 19$ (solid line) and 
$14$ (dashed line) both for $u_{\rm min} \simeq -0.03$. The first derivatives
of $\Theta_{\PM}$ and $\Theta_{\PP}$ diverge logarithmically for
$u\rightarrow 0$. 
The thin lines for $u>1$ indicate the asymptotic 
behaviors of $\Theta(u \gg 1)$ given in Eq.~\reff{eq:sfasy} with the numerical
values of the coefficients $A_\pm$ indicated from top to bottom for the
corresponding curves. For $\PP$ boundary conditions the asymptotic
expressions are indistinguishable from $\Theta_{\PP}(u)$
for $u\gtrsim 4$. %
}
\label{fig:DerjPot}
\end{figure}

\begin{figure}
\centering\includegraphics[width=\columnwidth]{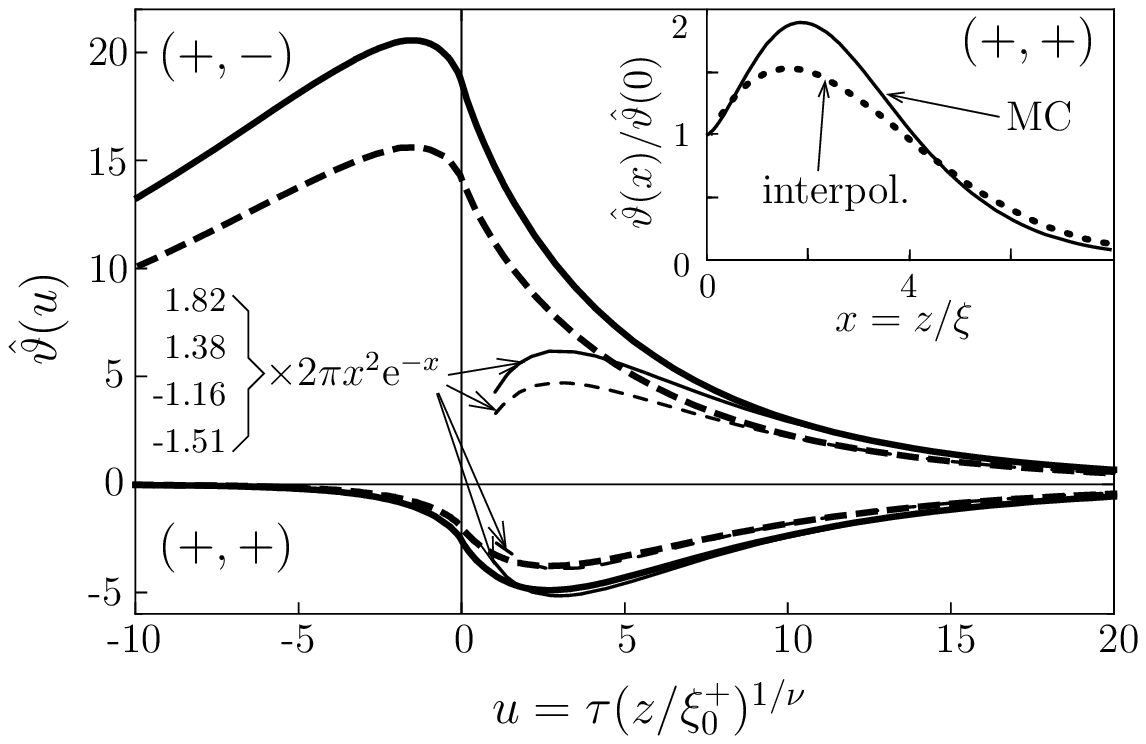}
\caption{Scaling functions  $\hat\sf_\PP$  and $\hat\sf_\PM$ of the Casimir
force $F_C$ [see Eq.~\reff{eq:scalF}] for
$\PP$ and $\PM$ BC, respectively, within the Derjaguin approximation and 
for the three-dimensional Ising universality class, as functions of $u = \tau 
(z/\xi_0^+)^{1/\nu}$ with $\nu \simeq 0.630$. 
The thick 
solid and dashed lines have been obtained via Eq.~\reff{eq:scalingfuncF} 
on the basis of the  Monte Carlo
estimates for $\sf_{\PP,\PM}$ presented in
Refs.~\protect{\cite{vas-07,vas-08}}, indicated by $(i)$ and $(ii)$,
respectively, in Figs.~9 and 10 of Ref.~\cite{vas-08}.  
$\hat\sf_{\PP}$ attains its minimum value $\hat\sf_{\PP}^{\rm (min)}
\simeq -4.9$ (solid line) and $-3.8$ (dashed line) both for 
$u_{\rm min} \simeq 2.6$, whereas $\hat\sf_{\PM}$ attains
its maximum value $\hat\sf_{\PM}^{\rm (max)} \simeq 21$ (solid line) and $16$
(dashed line) both for $u_{\rm min} \simeq -1.5$. 
The second derivatives of $\hat\sf_\PP(u)$ and  $\hat\sf_\PM(u)$ diverge
logarithmically for $u\rightarrow 0$. 
The thin lines for $u>1$ indicate the asymptotic 
behaviors $\hat\sf(u \gg 1)$ given in Eq.~\reff{eq:sfasy} with the numerical
values of the coefficients $A_\pm$ indicated from top to bottom for the
corresponding curves. For $\PP$ boundary conditions the asymptotic
expressions are indistinguishable from $\hat\sf_{\PP}(u)$
for $u\gtrsim 5$.
In the inset we compare the estimate for   
$\hat\sf_\PP(x)/\hat\sf_\PP(0)$
as a function of $x = z/\xi$ based on the Monte Carlo data of
Refs.~\cite{vas-07,vas-08} (solid line, MC) with the one presented
in Ref.~\cite{colloids1a} and obtained by interpolating linearly and
pointwise the exactly known film scaling functions in $d=2$ and $d=4$ in order
to obtain an estimate for $d=3$ (dotted line, interpol.). 
The Monte Carlo estimate for this ratio is the same for both data sets $(i)$
and $(ii)$ in Ref.~\cite{vas-08}.
}
\label{fig:DerjFor}
\end{figure}

For the universality class of the three-dimensional Ising model, the scaling
functions $\sf$ for the Casimir force in the film geometry -- which enter
into Eq.~\reff{eq:scalingfunc} -- have been determined 
in Refs.~\cite{vas-07,vas-08} for $\PP$
and $\PM$ BC (or, equivalently, $\MM$ and $\MP$ BC) by Monte Carlo
simulations.
Due to the presence of strong corrections to scaling, the \emph{amplitudes}
of the corresponding numerical estimates for $\sf_\PP(x)$ and $\sf_\PM(x)$ 
are affected by a systematic uncertainty of about 20\%~\cite{vas-07,vas-08}. 
The numerical data presented in Refs.~\cite{vas-07,vas-08} are very well
fitted by certain analytic ans\"atze (at least in the range of scaling variable
which has been investigated numerically) 
which, in turn, can be used in order to calculate the corresponding scaling
functions $\Sf_\PP$ and $\Sf_\PM$ for the potential (Fig.~\ref{fig:DerjPot}, 
see also Fig.~2(d) in Ref.~\cite{nature}) as well as $\hat\sf_\PP$ and
$\hat\sf_\PM$ for the force (Fig.~\ref{fig:DerjFor}).
The simulation data for the film scaling functions 
$\sf_\PP(x)$ and $\sf_\PM(x)$ can actually be fitted even by functions of
various \emph{shapes} (the asymptotic behavior of which for large $|x|$ is,
however, fixed, see further below). 
This leads to different estimates of the scaling functions 
outside the
range of the scaling variable for which the Monte Carlo data are currently
available. This results also in different estimates of 
$\Sf_\PP$, $\Sf_\PM$, $\hat\sf_\PP$, and $\hat\sf_\PM$ 
obtained from $\sf_\PP(x)$ and $\sf_\PM(x)$ via
Eqs.~\reff{eq:scalingfunc} and~\reff{eq:scalingfuncF}.
However, the uncertainty of the estimates for the shapes 
is negligible compared to the
inherent systematic uncertainty associated with the amplitudes of $\sf_\PP(x)$
and $\sf_\PM(x)$. For a detailed discussion of these issues we refer to
Ref.~\cite{vas-08}.
The critical Casimir force $f_\Cas(T,L)$ between two planar walls [see
  Eq.~\reff{eq:scf}] with symmetry-breaking boundary conditions 
is expected to vary as $\exp(-L/\xi)$ as a
  function of $L\gg\xi$ for  $\tau>0$ (see, e.g.,
  Ref.~\cite{Tr-09} and in particular the footnote 
Ref.~[23] therein). Accordingly, 
$\vartheta_{(+,\pm)}(x \gg 1)= A_\pm \, x^3 \rme^{-x}$ and from
  Eqs.~\reff{eq:scalingfuncF} and~\reff{eq:scalingfunc} one finds
\be
\begin{split}
\hat\sf_{(+,\pm)}(x\gg 1) &= 2\pi A_\pm\, x^2\,\rme^{-x}\quad\mbox{and}\\
\Sf_{(+,\pm)}(x\gg 1) &= 2\pi A_\pm\, x\,\rme^{-x},
\end{split}
\label{eq:sfasy}
\ee
for the critical Casimir force and potential, respectively, in the
sphere-plate geometry. The analysis of the Monte Carlo data presented in 
Figs.~9 and 10 of Ref.~\cite{vas-08} yields $A_+^{(i)}=-1.51(2)$ and
$A_-^{(i)} = 1.82(2)$, respectively, for the data sets therein 
indicated as $(i)$
  whereas it yields  $A_+^{(ii)}=-1.16(2)$ and $A_-^{(ii)} = 1.38(2)$ for the
  corresponding 
data set $(ii)$. (We recall here that the data sets $(i)$ and $(ii)$ turn
  out to be proportional to each other, see Refs.~\cite{vas-07,vas-08} for
  details.)

Figure~\ref{fig:DerjFor} shows that the critical Casimir force for the
sphere-plate geometry exhibits 
the same qualitative features as in the film
geometry: For $\PP$ [$\PM$] BC the force is attractive (repulsive) and attains
its maximum strength for $\tau>0$ ($\tau<0$), corresponding to the one-phase 
(two-phase) region. For fixed values of the scaling variable, 
the strength of the repulsive force for $\PM$ BC is larger than the one
of the attractive force in the case of $\PP$ BC.
The inset of Fig.~\ref{fig:DerjFor} compares the estimate for the 
scaling function $\hat\sf_{\PP}(x=L/\xi)$ --- up to its normalization
$\hat\sf_{\PP}(0)$ --- based on the Monte Carlo data of
Refs.~\cite{vas-07,vas-08} (solid line) with the early estimate of
Ref.~\cite{colloids1a}, which is based on the pointwise and linear
interpolation between the exactly known film scaling functions in $d=2$ and
$d=4$, such as to obtain an estimate of $\hat\sf_{\PP}(x=L/\xi)$ for $d=3$
(dashed line). Although this latter estimate captures correctly some
qualitative features of the universal scaling function
$\hat\sf_{\PP}(x=L/\xi)$, it fails to be quantitatively accurate, as the
comparison with the Monte Carlo estimate reveals. The same consideration
applies to the corresponding estimates for $\Sf_{\PP}$.

%
Equations~\reff{eq:PhiDgen} and~\reff{eq:scalingfunc}, together with
Fig.~\ref{fig:DerjPot}, form the theoretical basis for the interpretation of
the experimental results for the effective interaction potential between 
a spherical colloidal particle and a planar wall,
immersed into a binary liquid mixture at its critical
composition and near its consolute point.

%
%
\subsubsection{Deviations from strong adsorption}

The theoretical analyses 
presented above and in Refs.~\cite{vas-07,vas-08,nature}
assume that the confining surfaces are characterized by a sufficiently strong
preferential adsorption for one of the two components of the mixture,
corresponding to $(+)$ or $(-)$ fixed-point boundary conditions in the sense
of renormalization-group theory~\cite{Binder83,diehl:86:0}. 
Within the coarse-grained field-theoretical
description of the binary mixture close to a boundary ${\mathcal
B}$ in terms of the order parameter $\op$~\cite{Binder83,diehl:86:0}, the
preferential adsorption is accounted for by a surface contribution $-h_s
\int_{\mathcal B}\rmd S \, \op({\bf x}\in {\mathcal B})$ to the effective free
energy of the system, where the ``surface field'' 
$h_s$ summarily quantifies the strength of the preferential
adsorption. Indeed, $h_s > 0$ [$h_s<0$] favors $\op > 0$ [$\op <
  0$] at the boundary ${\mathcal B}$ so that, for $|h_s|$ 
large enough,
$|\phi(z)| \propto z^{-\beta/\nu}$  at normal distances
$z\rightarrow 0$ (but still large on molecular scales) 
from ${\mathcal B}$~\cite{diehl:97}. 
The  $(+)$  and $(-)$ boundary conditions correspond to the limits   $h_s
\rightarrow +\infty$ and $-\infty$, respectively, 
of strong preferential adsorption.
Within this coarse-grained description the gross features of the 
relation between $h_s$ and the material properties of the wall and the
mixture can be inferred from the behavior of experimentally accessible
quantities such as critical adsorption profiles or excess adsorption 
(see, e.g., Refs.~\cite{SWF-85,FD-95}). 
For a weak adsorption preference, the corresponding $h_s$ might be so small
that upon approaching the critical point one effectively 
observes a crossover in the kind of boundary
condition imposed on the order parameter.
The critical Casimir force reflects such~\cite{M-08} or related~\cite{SD-08}
crossover behaviors; in the 
film geometry, depending on the film thickness, the
force can even change sign~\cite{M-08,SD-08}.  
On the basis of scaling arguments one expects that for moderate adsorption
preferences the scaling function in Eq.~\reff{eq:scf} additionally depends on
the dimensionless scaling variables $y_{s,i} \equiv
a_ih_{s,i}L^{\Delta_1/\nu}$, $i=1,2$, where $h_{s,1}$, $h_{s,2}$ are the
effective surface fields at the two confining surfaces, $a_i>0$ are
corresponding non-universal constants, and $\Delta_1\simeq 0.46$ is the
so-called surface crossover exponent at the so-called 
ordinary surface transition~\cite{co-ord-norm,diehl:86:0}. 
One can associate a length scale $\ell_i\equiv
(a_i|h_{s,i}|)^{-\nu/\Delta_1}$ with each surface field, 
such that the theoretical predictions discussed before are valid for 
$L \gg \ell_i$, i.e., $y_{s,i}\rightarrow \pm\infty$, whereas corrections
depending on $\ell_i/L$ are expected to be relevant for $L\simeq \ell_i$.
For $\ell_i \gg L$, instead, 
the preferential adsorption of the wall $i$ is so weak that a
crossover occurs
towards boundary conditions which preserve the $\op \mapsto -\op$
symmetry and there appears to be 
no effective enhancement of the order parameter
upon approaching the wall. 
Heuristically, the length scales $\ell_i$ can be interpreted as extrapolation
lengths $z_{{\rm ex},i} \propto \ell_i$ in the sense that for 
small enough $\ell_i \neq 0$ the order
parameter profile behaves as $|\phi(z\rightarrow 0)| \sim
(z+z_{{\rm ex},i})^{-\beta/\nu}$~\cite{Binder83,CA-83,SDL-94} 
upon approaching the
wall $i$. 
Within the concept of an extrapolation length 
the effects  of a physical
wall with a moderate preferential adsorption (which implies $\ell_i \neq 0$)
on the order parameter are equivalent to those of a fictitious wall with strong
preferential adsorption (which means $\ell_i=0$) displaced by a distance 
$-z_{{\rm ex},i}$ from
the physical wall. Although 
this picture is consistent only within mean-field
theory~\cite{Binder83,CA-83} it turns out to be useful for the interpretation
of experimental results~\cite{SWF-85} and simulation data~\cite{SDL-94} as an
effective means to take into account corrections to the leading critical
behavior.
Assuming that this carries over to the critical Casimir forces, a film of
thickness $L$ and moderate adsorption at the confining surfaces is expected
to be
equivalent to a film with strong adsorption and thickness 
$z_{{\rm ex},1}+L+z_{{\rm ex},2} > L$. On the same footing, a sphere of
radius $R$ and a plate at a surface-to-surface distance $z$, both with
moderate preferential adsorption, should behave as a sphere of smaller radius
$R-z_{{\rm ex},{\rm sph}}$ and a plate at a distance $z_{{\rm ex},{\rm pl}} +
z + z_{{\rm ex},{\rm sph}}> z$, both with strong preferential
adsorption. 
We anticipate here that the interpretation of the experimental data presented
in Sec.~\ref{sec:res:cc} does not require to account for the effect
described above, even though we cannot exclude the possibility that 
such corrections might become detectable upon comparison 
with theoretical data with a smaller systematic
uncertainty than the ones considered here.

\subsection{Noncritical composition}

\label{sec:th:nc}

\subsubsection{General properties}
\label{subsubsec:GP}
In this section we consider   thermodynamic paths approaching 
the critical point from the one-phase region by varying
the temperature at fixed {\it off-critical} 
compositions, e.g., $c_A\ne c^c_{A}$, where $c_A$ is the concentration 
of the $A$ component of a mixture. For systems with a lower consolute point these paths lie below
the upwards bent  phase boundary of first-order phase transitions
 in the temperature-composition  $(T,c)$
parameter space (see, c.f., the vertical paths in Fig.~\ref{fig:WL-PD}(b)).
 Performing experiments along such paths
 is  another  useful and interesting 
 probe of the  critical  Casimir force, because 
 the corresponding Casimir scaling function acquires an additional
  scaling variable
$\Sigma=\textrm{sgn}(h)L/l_h$, where
  $l_h=l_0|(c_A-c^c_{A})/c^c_{A}|^{-\nu/\beta}$ and 
$l_0$ is a nonuniversal amplitude.
The bulk field $h$ is  proportional to the difference
$(\mu_A-\mu_B)-(\mu_A-\mu_B)_c$ of the chemical potentials of the two
components of a binary liquid mixture. If this  difference is nonzero one has
$c_A\ne c^c_{A}$ for species $A$ in the bulk.
The nonuniversal amplitude $l_0$ can be 
determined from the corresponding correlation length $l_h$ 
which is experimentally accessible by measuring the
scattering structure factor for various concentrations $c_A > c_A^c$ 
at $T=T_c$. This nonuniversal amplitude is actually related to the two
independent nonuniversal amplitudes $\xi_0^+$ and $C_\phi$ 
(see the discussion below Eq.~\reff{eq:xidiv})
by the expression \cite{colloids1b}:
\begin{equation}
\label{eq:ampll_0}
l_0=\xi_0^+\left(\frac{C_\phi}{c^c_{A}}\right)^{\nu/\beta}
\left(\frac{Q_2}{\delta R_{\chi}}\right)^{\nu/\gamma}
\end{equation}
where $\delta$ and $\gamma$ are standard bulk critical exponents and 
$Q_2$ and $R_{\chi}$ are universal amplitude ratios \cite{tarko,PV,Priv}
leading to $[Q_2/(\delta R_{\chi})]^{\nu/\gamma}\approx 0.38$
in $d=3$. 

So far,  for the sphere-plane geometry of the present experiment there are no  theoretical results available for the critical Casimir force 
for thermodynamic states which lie off the bulk critical composition. 
However, based on the theoretical analysis of the 
critical Casimir force for  films \cite{DME,colloids1b} and sphere-sphere geometries
\cite{colloids1b},
we expect that   along suitably chosen  paths of fixed
off-critical compositions the critical Casimir force 
is strongly influenced by capillary bridging transitions.
Moreover, if the bulk field $h$ is nonzero
 $\PP$ and $\MM$ BC are no longer equivalent.

\begin{figure}
\centering\includegraphics[width=0.75\columnwidth]{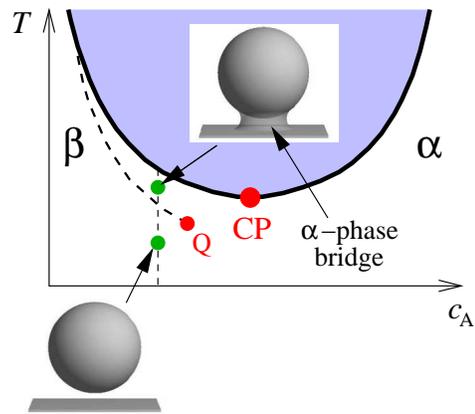}
\caption{%
Schematic phase diagram of a binary liquid mixture with a lower
demixing transition point in terms of temperature $T$ and concentration $c_A$
of the $A$ species.  The solid curve encloses the two-phase region separating
via first-order phase transitions the $\alpha$ and $\beta$ phases rich in $A$
and $B$ species, respectively, terminating at the critical point  CP.
The dashed line indicates  the first-order bridging  phase transition
which  occurs
if a fluid mixture is confined between a planar wall and a sphere of radius
$R$ and at distance $L$ possessing  the same
adsorption preference, here for the $\alpha$ phase.
The bridging transition ends at the critical point $Q$ and separates a
region in the bulk phase diagram in which a phase preferable  by walls
condenses  and forms a bridge connecting the wall and the sphere,  from the
region in the bulk phase diagram where such
a bridge is absent. Although the bridging transition is a (quasi-) first-order
phase transition, in the {\it bulk} phase diagram it is described by a line
instead of a coexistence region, because
it is an interfacial phase transition. }
\label{fig:phasediagram}
\end{figure}

\subsubsection{Bridging transition}
\label{subsubsec:GP}
A bridging transition is the 
analogue of capillary condensation \cite{evans} for  geometries 
in which  one or both surfaces are non-planar.
(However, there is a conceptual difference. Whereas capillary condensation corresponds 
to an actual shift of the bulk phase diagram, bridging transitions are interfacial phase transitions which leave the bulk phase diagram unchanged but can be described as if effectively
the bulk phase boundary of first-order phase transitions is 
shifted~\cite{BD,BBD}.)
 It occurs 
at  temperatures for  which two phases may exist, i.e., 
for $T$ above $T_c$ in the case of a binary liquid mixture with a lower 
consolute point, and it depends on the adsorption properties of the
surfaces.
If, say, both  surfaces favor the $\alpha$ phase rich in species A
over the $\beta$ phase rich in species B,  one expects 
the $\alpha$  phase to form a bridge  between the surfaces for some chemical
potential $\mu_A$ of species A such that
$\mu_A<\mu^{co}_A$, where $\mu^{co}_A$ is the value corresponding to bulk
coexistence. Alternatively, this occurs 
at a concentration (mole fraction) $c_A<c_A^{co}$
slightly smaller than its value $c^{co}_A$ at bulk coexistence.
If the surfaces favor the $\beta$ phase,  the $\beta$ phase 
fills the gap between the surfaces forming a bridge for
$\mu_A > \mu^{co}_A$, i.e., the phase separation line for this morphological transition occurs on the other side of the bulk phase diagram, i.e., for  $c_A>c^{co}_A$
 (Fig.~\ref{fig:phasediagram}).

 Bridging may occur  in the presence 
of thin wetting layers on both surfaces, i.e., in the partial wetting 
regimes of the two individual  surfaces
 \cite{dobbs,andrienko,shinto}, or if  one or both surfaces 
are covered by a thick wetting film~\cite{BBD}.
Such bridge formation 
 may be relevant for colloid aggregation or flocculation 
of the particles \cite{beysens1,beysens2} 
(for a  summary of the corresponding experimental
and theoretical work on these phenomena see Refs.~\cite{BD,BBD}).
 For the 
 sphere~--~planar wall
geometry relevant  for the present experimental situation, theoretical studies
\cite{andrienko} predict  that the bridging transition can occur  in the
presence of  thin wetting layers coating  both surfaces.
It  is a first-order phase transition 
and ends at a critical point. (Actually, these bridging transitions are only
quasi-phase transitions, because they involve, strictly speaking only a
zero-dimensional volume \cite{BD,BBD}).
For a fixed distance between 
the wall and the sphere and fixed chemical potential, the position of this
critical point is determined by the relation 
$\xi \simeq R$, where $R$ is the radius of the  sphere  (see
Fig.~\ref{fig:phasediagram}).  For small sphere radii the bridge configuration
is unstable, even for very small sphere-plane separations.
On the other hand, bridging transitions are  not possible  for large
sphere-plane separations, even if the sphere radii are  large.
The fluid-mediated solvation  force between the surfaces
 is very weak in the absence of the bridge and  
it is attractive  and long-ranged if  the capillary bridge is present.
Moreover, for $R/\xi$ small its strength is proportional to the sphere-wall
separation~\cite{andrienko,shinto}, contrary to 
 the case of two flat substrates~\cite{evans}
or to the sphere-sphere geometry~\cite{BBD}. 

\subsubsection{Critical Casimir forces for noncritical compositions}
\label{subsub:CC}

For temperatures closer to the critical temperature the solvation force 
acquires a universal contribution due to the critical fluctuation of the intervening
fluid which turns into  the critical Casimir force.
For a one-component fluid near gas-liquid  coexistence 
$\mu=\mu_0(T)$ and confined between   parallel plates it has been shown  \cite{DME}
that at temperatures
near the critical temperature $T_c$ a small 
bulk-like field
 $h\sim \Delta \mu = \mu-\mu_0(T)<0$, which favors the gas  phase, leads
to residual condensation and consequently to a critical Casimir force which, at the same 
large wall separation, is much more attractive than the one
 found exactly at the critical point.
The same scenario is expected to apply  to  binary liquid mixtures, i.e., the Casimir force
is expected to be much more attractive  for compositions 
slightly away from the critical composition on that side of the bulk phase diagram 
which corresponds to the bulk phase disfavored by the confining walls. This has been studied in detail
in Ref.~\cite{colloids1b} by using the standard
 field-theoretic model within  mean-field approximation. 
These studies of the parallel plate geometry have been extended to the case of two spherical 
particles  of radius $R$ at a finite distance $L$ \cite{colloids1b}. 
The numerical results for the effective pair potential, as well as the
results obtained by using the knowledge of the force between
  parallel plates and then by applying the Derjaguin approximation,
 valid for $L\ll R$, show that at $T=T_c$
 the dependence of the Casimir force on the composition  
 exhibits a pronounced 
maximum at a noncritical composition.
One expects that such a shift of the force maximum to noncritical compositions
results from  the residual capillary bridging
and that the direction of the shift relative to the critical composition depends on the boundary conditions. 
If the surfaces prefer the $\alpha$ phase rich in species A, 
 by varying the temperature at fixed  
off-critical composition $c_A$, one observes
that for   small deviations $|c_A-c^c_{A}|\ll c^c_{A}$,
 the position of the maximum of the Casimir  force as function of temperature  is almost unchanged,  while
the absolute value of the maximal force increases considerably by moving away
from  $c^c_A$ to compositions $c_A<c^c_A$. The overall temperature  variation is, however, similar to that
at $c^c_{A}$, provided one stays sufficiently close to the critical composition.
For compositions $c_A$ slightly larger than the critical composition, $c_A >c^c_A$,
the critical Casimir force as a function of temperature is expected to behave 
in a similar way as for $c_A<c^c_A$, but the amplitude of the force maximum should be much weaker
and should decrease for increasing $c_A$.
At compositions further away from $c^c_A$, i.e., off the critical regime, 
due to the small bulk correlation length the Casimir force is vanishingly small unless the 
aforementioned bridging transition is reached by varying  the temperature.

The case of a sphere against a planar wall has not been 
studied theoretically. However, we expect  a similar behavior of the effective
forces as the one for two spheres.

\section{Experiment} \label{sec:exp}

\subsection{The method: Total Internal Reflection Microscopy}
\label{sec:exp_TIRM}

\subsubsection{Basic principles of TIRM}

Total internal reflection
microscopy (TIRM) is a
technique which allows one to determine the
potential $\Phi$ of effective forces acting on a single 
colloidal particle suspended in a
liquid close to a planar substrate, with a force resolution down to
the order of femto-Newton. 
The potential $\Phi$ is obtained
from the probability distribution to find the surface of the particle at
height $z$ above the substrate, which
is determined by monitoring the Brownian motion of the particle in the 
direction perpendicular to the substrate.
In TIRM measurements this is achieved by creating an
evanescent light field  at the substrate-liquid interface 
which penetrates into the liquid.
The intensity of the evanescent field 
varies strongly with the distance from the substrate. A single 
colloidal particle
scatters light if it is illuminated by such a field. From this scattered
intensity it is possible to deduce the position of the particle in the
evanescent field, i.e., to determine $z$ and its time
dependence~\cite{Walz-97,Prieve-99}. 
%

\begin{figure}
\centering\includegraphics[width=0.7\columnwidth]{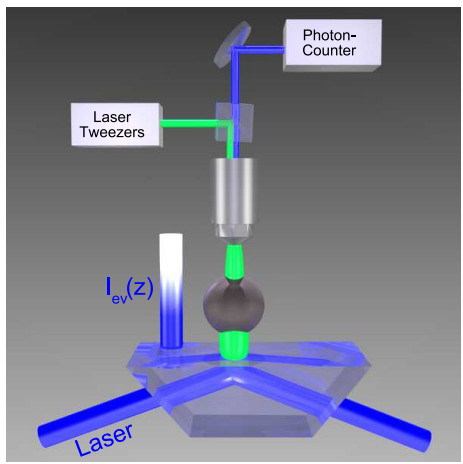}
\caption{%
Data acquisition system (see main text for details). The green laser light
generated by the optical tweezers is deflected by a double prism into the
microscope objective (in the figure represented as a grey vertical cylinder
above the spherical particle) 
and it provides an optical potential which confines
the spherical colloid laterally. 
The blue light which is scattered by this particle out of the
evanescent field of intensity $I_{\rm ev}(z)$ is collected by the same
microscope objective, focused and then optically directed into 
the photon counter via a combination of prisms and mirrors (schematically
represented in the upper part of the figure).} 
\label{fig:data_acq}
\end{figure}
%
%
The basic experimental setup is presented in
Fig.~\ref{fig:data_acq}. A p-polarized laser beam ($\lambda =
473$nm, $P=2\,\mbox{mW}$) is directed from below 
onto the interface between the bottom of
a silica glass cell (a cuvette with a chamber to accommodate 
a fluid film of thickness $200\,\mu\mbox{m}$)
and the liquid containing the
colloidal particle. The illumination angle $\theta_i$ (formed with the
substrate normal) is larger
than the critical angle $\theta_c$ of total internal reflection.
Due to total internal reflection, 
an evanescent wave penetrates into
the medium with lower refractive index, here the liquid,
and its intensity $I_{\rm ev}(z)$ decays exponentially as a
function of the distance $z$ from the glass-liquid interface:
\be
I_{\rm ev}(z) = I_{\rm ev}(0)\rme^{-\zeta z} \;.
\label{eq:Iev}
\ee
The decay constant $\zeta$ defines the 
{\it penetration depth} $\zeta^{-1}$, which is given by~\cite{Prieve-99} 
\be  
\zeta^{-1} = \frac{\lambda}{4 \pi  \sqrt{n_{\rm glass}^2
\sin^2\theta_{i} - n_{\rm liq}^2}} \,, 
\label{eq:zeta}
\ee
where $\lambda $ is the wavelength of the illuminating
laser beam in vacuum,  
and $n_{\rm glass}$ and $n_{\rm liq}$ are the refractive indices of the
glass and the liquid, respectively. 
In our experiment (see, c.f., Sec.~\ref{sec:exp_bm}) the critical binary 
mixture
(liquid) has $n_{\rm liq} = 1.384$ whereas the silica glass
(substrate) has 
$n_{\rm glass} = 1.464$ ($>n_{\rm liq}$), resulting in $\theta_c
\simeq 71^\circ$.
A colloidal particle with a refractive index $n_{\rm coll} > n_{\rm liq}$
(in our experiment the polystyrene colloids have $n_{\rm coll}= 1.59$)
at a distance $z$ away from the surface 
scatters light from the evanescent field.
Within the well established data evaluation model
for TIRM intensity, the light scattered by the 
particle has an intensity $I_{\rm sc}$ 
which is proportional to $I_{\rm
sc} \propto I_{\rm ev}(z)$~\cite{Prieve-99} and therefore depends on the
distance $z$.
Care has to be taken in choosing parameters for the
penetration depth and the polarization of the illuminating laser
beam in order to avoid optical distortions due to multiple reflections
between the particle and the substrate, which would spoil the linear
relation between $I_{\rm sc}$ and $I_{\rm ev}(z)$.
In this respect, safe parameter regions 
are known to be small penetration depths $\zeta < 250$nm and
p-polarized illumination as used in the present experiment~\cite{helden-06,
hertlein-08}.
As a result of this relation,
the scattered light intensity $I_{sc}$ exhibits an
exponential dependence on the particle-wall distance with exactly
the same decay constant $\zeta^{-1}$ ($\zeta^{-1} = 200\pm 2\,$nm in
our experiment) 
as the evanescent field intensity $I_{\rm ev}$:
\be 
I_{\rm sc}(z) = I_0 \; \rme^{-\zeta z}\,, 
\label{eq:TIRM} 
\ee
where the scattered intensity  $I_0$ at contact $z=0$ depends on
the laser intensity, the combination of refractive indices, and the 
penetration depth. 
As will be discussed below, the knowledge of $I_0$ is important 
to determine the particle-substrate distance from the scattered intensity
$I_{\rm sc}$. In principle $I_0$ could be measured by the so-called sticking
method~\cite{Prieve-99} according to which 
the particle is stuck on the substrate 
due to the addition of salt to the liquid in such a way as to suppress 
the electrostatic stabilization which normally repels the particle from the
substrate. 
However, in the system we are interested in this is not practicable given 
the large concentration of salt ($> 6\,$mM) required to force the particle
to stick to the surface and the compact design of the experimental  
cell which limits the access to the sample.
Instead, as described further below, 
we circumvent this problem by using a hydrodynamic
method~\cite{bevan00} for the absolute determination of the particle-substrate
distance.

In Sec.~\ref{sec:th} we mentioned that, upon approaching the critical
point of the binary liquid mixture, critical adsorption profiles form near the
surfaces of the substrate and of the colloid. 
These concentration profiles induce a spatial variation of the refractive
index, which deviates from the assumed steplike variation underlying
Eqs.~\reff{eq:Iev} and~\reff{eq:zeta} (see, e.g.,
Ref.~\cite{crit-ads}). Deviations from the functional form given by 
Eq.~\reff{eq:Iev}
are pronounced if the correlation length $\xi$ becomes comparable with the
wavelength $\lambda$ of the laser light, which is not the case for the
experimental data obtained here, for which $\lambda = 473\,$nm and
 $\xi \lesssim 100\,$nm (see, c.f., Figs.~\ref{fig:fitximm},
\ref{fig:fitxipm}, and \ref{fig:fitximp}).

\begin{figure*}
\centering
\begin{tabular}{ccc}
\includegraphics[width=5cm]{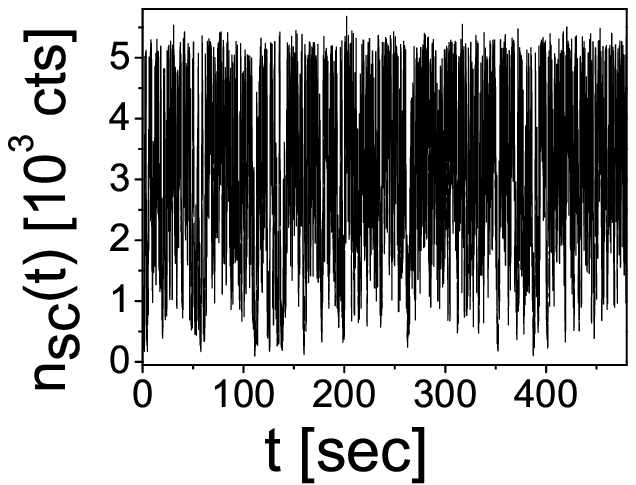} &
\includegraphics[width=5cm]{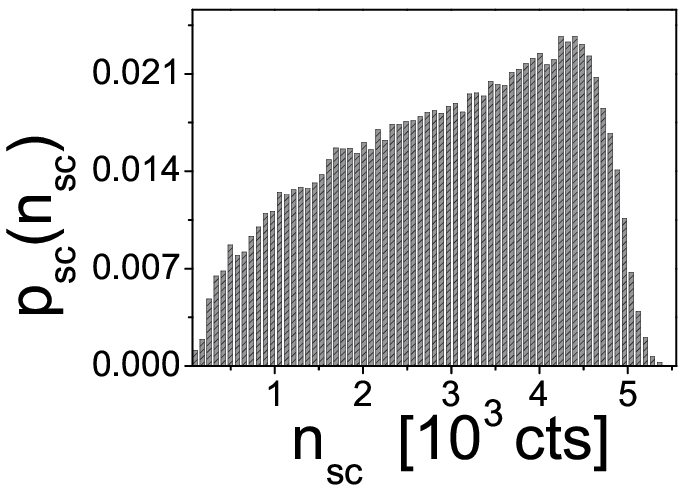} &
\includegraphics[width=5cm]{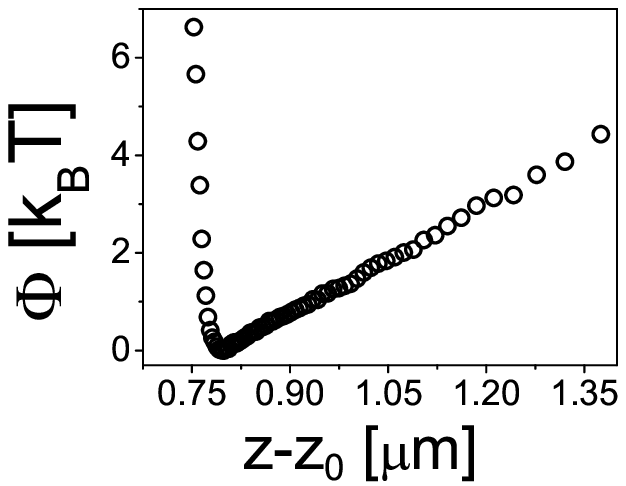} \\
(a)&(b)&(c)
\end{tabular}
\caption{%
Data analysis: 
(a) Raw data for the total number of scattered photons $n_{\rm sc}(t)$
detected at a certain time $t$ within a time interval $\Delta t =
1\,\mbox{ms}$ from the single photon counter as
a function of time, taken for a sampling time $t_{\rm samp}\simeq
15\,$min. 
(b) Histogram calculated from this time series yielding the distribution
function $p_{\rm sc}(n_{\rm sc})$ of the number $n_{\rm sc}$ of scattered
photons. 
(c) The knowledge of the relation [see
Eq.~(\protect{\ref{eq:TIRM}})] between the scattered intensity $I_{\rm
  sc}\simeq n_{\rm sc}/\Delta t$ and
the position $z$ of the colloid in the evanescent field allows one 
to determine the probability distribution 
$p_z(z)=\zeta n_{\rm sc}(z) p_{\rm sc}(n_{\rm sc}(z))$ for the
particle-substrate distance $z$ 
from $p_{\rm sc}(n_{\rm sc})$ and therewith by inversion of the Boltzmann
factor the interaction potential. Further details are given 
in the main text.
}
\label{fig:data_analy}
\end{figure*}

In a typical TIRM measurement run, 
the vertically scattered intensity  
$I_{\rm sc}(t) = I_{\rm sc}(z(t))$ (photons/s) 
is recorded by a photomultiplier connected to a
single photon counter [see Fig.~\ref{fig:data_acq}]
which counts the
total \emph{n}umber of scattered photons [see
  Fig.~\ref{fig:data_analy}(a)]
\be
n_{\rm sc}(t) \equiv \int_t^{t + \Delta t}\!\!\rmd t' I_{\rm sc}(t') \simeq
I_{\rm sc}(t)\Delta t
\label{eq:nsc}
\ee
detected within a time interval
$\Delta t = 1\,$ms~\cite{footnote_nvsI}. 
The value of $n_{\rm sc}(t)$ is then acquired with a
frequency $f_{\rm samp}=250\,$Hz for a total duration $t_{\rm samp}\simeq
15\,$min. 
The resulting set of data is then analyzed as
described below in Sec.~\ref{sec:exp_TIRM_da}.
Consecutive intensity data $I_{\rm sc}(t)$, i.e., $n_{\rm sc}(t)$ 
acquired with
a larger frequency $f_{\rm samp}$ turn out to be strongly correlated in
time. Accordingly, their acquisition does not contribute to the reduction of
the statistical errors affecting the final estimate for the potential, as will
be  discussed in, 
c.f., Sec.~\ref{sss:ct}. This observation motivates our choice
$f_{\rm samp}=250\,$Hz.

In addition to the detection optics, an optical tweezer is
implemented in the TIRM setup~\cite{ash-70} in order to be able to 
control 
the lateral position of the particle. 
The tweezer is created by a
laser beam ($\lambda_{\rm tweezer} = 532\,$nm) incident on the
particle from the direction perpendicular to the substrate 
and focused by the microscope objective used also for
the detection (see Fig.~\ref{fig:data_acq}).
Via this tweezer it is possible to conveniently position the probe
particle within the measuring cell and to restrict its lateral diffusion to
a few microns so that the particle 
does not diffuse out of the field of view of the detection
system. In addition, the tweezer also exerts a 
light pressure~\cite{walz92} onto the 
particle, increasing significantly its effective weight (in the specific case
considered here from ca.~$1.1 \kB T_c/\mu\mbox{m}$ to ca.~$7 \kB
T_c/\mu\mbox{m}$, see, c.f., Sec.~\ref{subsec:IP} and 
Refs.~\cite{footnoteB,dataWL} 
for details).
In our experiment the tweezer is typically
operated at a low power of $P \simeq 2\,$mW, but even at the largest power
($P_{\rm max} \simeq 25\,$mW)  we used to trap and move the particle 
no effects of local heating, such as the onset of 
phase separation in 
the liquid, were observed due to the laser of the tweezer.

\subsubsection{Data analysis}
\label{sec:exp_TIRM_da}

%

\begin{figure}
\centering
\includegraphics[width=0.85\columnwidth]{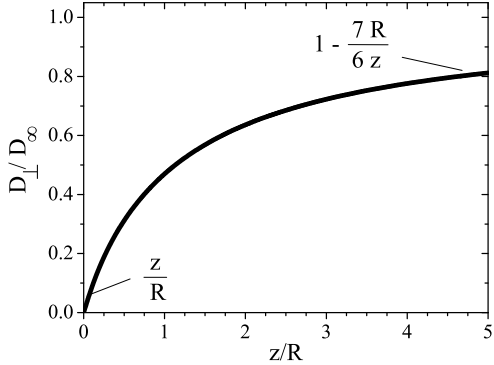}
\caption{Diffusion
coefficient $D_\bot$ of a spherical particle of radius $R$ moving 
perpendicular to a wall at a distance $z$~\cite{gold67,bevan00}. 
$D_\infty$ is the bulk diffusion coefficient. 
Note that for the present experimental conditions 
$z/R<1$ and thus this distance dependence is pronounced.}
\label{fig:Dsenkrecht}
\end{figure}

%
%
In order to determine the potential $\Phi$ of the effective 
forces acting on the
colloidal particle, one constructs a histogram out of the values of
$n_{\rm sc}(t)$ [see Eq.~\reff{eq:nsc}] 
recorded in the time interval $t_{\rm samp}$, in such a way as
to determine the probability distribution function $p_{\rm sc}(n_{\rm sc})$ 
for the particle
to scatter  $n_{\rm sc}=I_{\rm sc}\Delta t$ photons in a time 
interval $\Delta t$. Within the sampling time $t_{\rm samp}$ there are $t_{\rm
samp} f_{\rm samp} = N$ registration of counts. If $N^*$ is the number of
registrations which yield a certain count $n_{\rm sc}^*$, the probability
$p(n_{\rm sc}^*)$ of $n_{\rm sc}^*$ to occur is $N^*/N$.
By using $p_{\rm sc}(n_{\rm sc})\rmd n_{\rm sc} = p_z(z)\rmd z$ and 
Eq.~\reff{eq:TIRM}, this probability
distribution  $p_{\rm sc}(n_{\rm sc})$ can be transformed into the probability 
\be
p_z(z)=\zeta \,n_{\rm sc}(z)\, p_{\rm sc}(n_{\rm sc}(z))
\label{eq:pz}
\ee
for the particle-substrate distance $z$.
In turn, in thermal equilibrium at temperature $T$, 
the probability $p_z(z)$ is related to
the particle-wall interaction potential $\Phi(z)$ by the Boltzmann
factor 
\be 
p_z(z) = C \exp\left[-\Phi(z)/(\kB T)\right], 
\label{eq:BF}
\ee 
where $k_BT$ is the thermal energy and $C$ a normalization constant.
Equation~\reff{eq:BF} holds because, due to the high dilution of the colloidal
suspension, the single colloidal particle under observation does not interact
with other particles.
As a result, from the knowledge of
$p_{\rm sc}(n_{\rm sc})$ 
it is possible to determine $\Phi(z)$ up to an irrelevant
constant related to $C$ and to the overall normalization of 
$p_{\rm sc}(n_{\rm sc})$.
For each bin of the histogram of $p_{\rm sc}(n_{\rm sc})$,
the corresponding distance $z(n_{\rm sc})$ is 
calculated via inversion of Eq.~\reff{eq:TIRM}:
\be
z(n_{\rm sc}) = -\zeta^{-1} \ln [n_{\rm sc}/(I_0\Delta t)] 
=  z_{\rm exp}(n_{\rm sc}) - z_0
\label{eq:z}
\ee
where $z_{\rm exp}(n_{\rm sc}) = - \zeta^{-1} \ln n_{\rm sc}$ is given
in terms of 
experimentally accessible quantities (i.e., $n_{\rm sc}$ and $\zeta$).
This provides the position of the particle up to 
the constant $z_0 = - \zeta^{-1} \ln (I_0\Delta t)$ as the experimentally 
yet unknown position of the wall~\cite{footnote1} [$I_{\rm sc}(z=0)=I_0$]. 
In order to determine $z_0$ for all data 
sets, we have employed the
so-called 
hydrodynamic method~\cite{bevan00}, which is based on the fact that due to
hydrodynamic interactions the
diffusion coefficient $D$ of a colloidal particle at a distance $z$ 
from the wall strongly depends on  $z$.
Moreover, near a wall 
the diffusion coefficient becomes also spatially anisotropic
with the relevant value for TIRM measurements being $D_\bot$, 
which refers to the diffusion occurring in the direction perpendicular to the
wall. Its spatial dependence can be expressed as 
\be 
D_\bot= D_{\infty} f(z/R) 
\label{eq:Dperp}
\ee
where $D_\infty = \kB T/(6\pi \eta R)$ is the bulk diffusion
coefficient of a spherical particle of radius $R$ in a homogeneous fluid
with viscosity $\eta$ at temperature $T$. (For the water-lutidine mixture we
use in our experiments, the value of $\eta$ has been measured in
Ref.~\cite{clu99} as a function of temperature $T$ and composition, with $\eta
\simeq 2.09 \times 10^{-3}$Pa\,s at $T=31^\circ$C and at the 
critical composition.) 
The reduced mobility function $f(v)$ was calculated in Ref.~\cite{gold67} 
and can be well approximated by~\cite{bevan00}:
\be 
f(v)= \frac{6v^2 + 2v}{6v^2 + 9v + 2} \,.
\label{eq:diff} 
\ee
A plot of this theoretically predicted distance-dependent diffusion coefficient
$D_\bot$ is shown in Fig.~\ref{fig:Dsenkrecht}.
%
A well established method~\cite{bevan00} 
to determine the absolute particle-wall distance
is to calculate the  
apparent diffusion coefficient $D_{\rm app}$ which is the
weighted average of $D_\bot(z)$ over the distances sampled by the colloidal
particle, i.e., $D_{\rm app} =
\int_0^\infty\rmd z\, D_\bot(z) \rme^{-2\zeta z}p_z(z)/\int_0^\infty\rmd
z\,\rme^{-2\zeta z}p_z(z)$, where the exponential factors in the numerator and
denominator reflect the spatial dependence of $I^2_{\rm sc}(z)$ 
[see Eq.~\reff{eq:TIRM} and below].
This apparent diffusion coefficient 
can be experimentally determined from the initial slope of the autocorrelation
function $C(\delta t) = \langle n_{\rm sc}(t) n_{\rm sc}(t+\delta t) \rangle_t$
of the scattering intensity~\cite{bevan00}:
\be 
D_{\rm app}= - \frac{1}{\zeta^2} \frac{C'(0)}{C(0)}
\label{eq:Dapp} 
\ee
where the prime $'$ denotes the derivative with respect to $\delta t$. 
In order to determine $z_0$ one calculates the apparent diffusion coefficient
$D_{\rm app, calc}$ on the basis of $D_\bot(z)$ and of the experimentally
determined probability distribution $p_z$ which is given 
by the parametric plot of $\zeta n_{\rm sc} p_{\rm sc}(n_{\rm sc})$ as a
function of $z_{\rm exp}(n_{\rm sc})-\hat z_0$ upon varying $n_{\rm sc}$ 
[Eqs.~\reff{eq:pz} and~\reff{eq:z}], with $\hat z_0$ as the yet unknown
position of the wall, and  according to which the colloidal particle samples
distances.
In turn, the value $\hat z_0 = z_0$ can be determined by requiring that
$D_{\rm app}=D_{\rm app, calc}$. A detailed description of this procedure can
be found in Ref.~\cite{bevan00}. 
The uncertainty in the determination of $z_0$ via this method (see Appendix B
of Ref.~\cite{bevan00}) is
primarily determined by the uncertainties of the
particle radius [see Eqs.~\reff{eq:Dperp} and \reff{eq:diff}] 
and of the penetration depth 
$\zeta^{-1}$ [see Eq.~\reff{eq:Dapp}]. 
Considering the experimental parameters and errors of our measurements,  
the resulting uncertainty in the particle-substrate distance $z$ can 
be estimated to be $\pm 30\,$nm for all plots shown in the following.

We emphasize that it is sufficient to determine $z_0$ at a certain
temperature in order to fix it for all the measured potential curves at
different temperatures. Indeed the intensity $I_{\rm ev}(0)$ of the evanescent
field at the glass-liquid surface 
[as well as $\zeta^{-1}$, see Eq.\reff{eq:zeta}] depends on
temperature via the temperature dependence of the optical properties of the
glass and the liquid. In turn, this would imply a variation of the critical
angle $\theta_c$ with $T$, which was actually not observed within the range of
temperatures investigated here.  The intensity $I_0 \propto I_{\rm
  ev}(0)$, which determines $z_0$ and which is recorded by
the photomultiplier, is in principle affected by the
temperature-dependent background light scattering due to the critical
fluctuations within the mixture (critical opalescence). 
For the typical intensities involved in our
experiment and for the temperature range studied, 
the contribution of this background scattering is actually
negligible and, as a result, $z_0$ does not change significantly with
temperature. 
The hydrodynamic method, however, requires the knowledge of the
viscosity $\eta$ of the mixture, which depends on temperature and 
sharply increases upon approaching the critical point~\cite{clu99} due to
critical fluctuations. These fluctuations might in addition
modify the expression of $D_\perp(z)$. In order to reduce this influence
of critical fluctuations we have chosen $T_c - 3\,$K  
as the reference temperature for determining $z_0$, corresponding to
a temperature at 
which no critical Casimir forces could actually be detected in the interaction
potential.
%

\subsubsection{Interaction potentials}
\label{subsec:IP}

Under the influence of gravity, buoyancy, and the radiation field of the
optical tweezer as external forces, the total
potential $\Phi$ of the colloidal particle floating in the
binary liquid mixture, as determined via TIRM, 
is the sum of four contributions: 
\be
\Phi(z) = \Phi_0(z) + G_{\rm eff} z + \Phi_C(z) + \Phi_{\rm
offset}\,. 
\label{eq:potgen}
\ee 
In this expression $\Phi_0$ is the potential due to
the electrostatic interaction between the colloid and the wall and
due to dispersion forces acting on the colloid; it is typically
characterized by a short-ranged repulsion and a long-ranged attraction.
The combined action of gravity, buoyancy, and light pressure from the optical
tweezer is responsible for  the linear term $G_{\rm eff} z$ in
Eq.~\reff{eq:potgen}  (see, e.g., Ref.~\cite{walz92}). 
$\Phi_C(z)$ is the critical Casimir potential arising from the critical
fluctuations in the binary mixture. 
The last term $\Phi_{\rm offset}$ is an undetermined, spatially constant 
offset different for each measured potential which accounts for the 
potentially
different normalization constants of the distribution functions $p_I$ and
$p_z$. 
While the first two contributions are expected to depend mildly on the
temperature $T$ of the fluid, the third one should bear a clear signature of
the approach to the critical point.
These expectations are supported by the experimental findings reported in
Sec.~\ref{sec:res}. The typical values of $G_{\rm eff}$ for the measurements
presented in Sec.~\ref{sec:res} are $G_{\rm eff} \simeq 7.2 k_B T/\mu$m and
$G_{\rm eff} \simeq 10.0 k_B T/\mu$m for the colloids with diameters 
$2R=2.4\mu$m and $3.68\mu$m, respectively~\cite{footnoteB}.  
Far enough from the surface, $\Phi_0$ and $\Phi_C$ are negligible
compared to the linear term and therefore the typical potential
$\Phi(z)$ is characterized by a linear increase for $z$ large enough.
Accordingly, upon comparing  potentials determined experimentally 
at different temperatures, the corresponding additive constants $\Phi_{\rm
offset}$, which are left undetermined by the TIRM method, can be
fixed consistently such that the linearly increasing parts of the various 
$\Phi$ coincide. 
However, it may happen that at some temperatures the total potential $\Phi$
develops such a deep potential minimum that the colloid cannot escape from it 
and therefore the gravitational tail is not sampled. If this occurs the
shift of this potential by a constant cannot be fixed by comparison with
the potentials
measured at different temperatures. 
In order to highlight the interesting contributions to the potentials, the
term $G_{\rm eff} z$, common to all of them,  is
subtracted within each series of measurements and for all boundary
conditions. Accordingly, the remaining part of the potential 
--- displayed in the figures below --- decays to zero at large distances.

On the other hand, closer to the substrate, 
the (non-retarded) van-der-Waals forces contribute to
$\Phi_0(z)$ with a term (see, e.g., Ref.~\cite{vdW}, Tab. S.5.b)
\be
\Phi_{0,{\rm vdW}}(z) = - \frac{A}{6} \left[\frac{1}{\delta} + \frac{1}{2
  + \delta} - \ln(1+2/\delta)\right],
\label{eq:vdW}
\ee
where $A$ is the Hamaker constant and $\delta = z/R$. As $\delta$ increases,
this term crosses over from the behavior $\Phi_{0,{\rm vdW}}(z\ll R) \simeq
-(A/6) (R/z)$ to $\Phi_{0,{\rm vdW}}(z\gg R) \simeq
-(2A/9) (R/z)^3$. The dependence of $\Phi_{0,{\rm vdW}}(z\ll R)$ on $z$ is the
same as the one of $\Phi_C(z\ll \xi,R)$ (see Eq.~\reff{eq:PhiDgen}) and
therefore their relative magnitude is controlled by $|\Phi_{0,{\rm vdW}}(z\ll
R)/\Phi_C(z\ll \xi,R)| = [A/(\kB T)]1/[6\Theta(0)]\simeq 2.4/(6\Theta(0))$ for
a critical point at $T\simeq 300$ K and a typical value of the Hamaker constant
$A \simeq 10^{-20}$ J. Taking into account that $\Theta_{\PP}(0) \simeq 2.5$
and $\Theta_{\PM}(0)\simeq 15$ it is clear that in this range of distances
the critical 
Casimir potential typically dominates the (non-retarded) van-der-Waals
interaction. Note, however, that for small values of $z$ both of them become
negligible compared to the electrostatic repulsion. For larger values of $R
\ll z\ll \xi$, the Casimir force still dominates the
dispersion forces, as discussed in detail in 
Refs.~\cite{colloids1a,colloids1b}. 
However, for the present
experimental conditions, the distance $z$ is
comparable to the bulk correlation length $\xi$ and actually most of the data
refer to the case $\xi\lesssim z \lesssim R$, with the values of the
scaling variables ranging between $z/\xi \simeq 10$
and $z/R \simeq 0.6$ for distances at which the
corresponding potential is still experimentally 
detectable. Accordingly, with the above estimate for the potential ratio, 
one can conservatively
estimate $|\Phi_{0,{\rm vdW}}(z\sim 0.6
R)/\Phi_C(z\sim 10 \xi)| = [A/(\kB T)] (0.2/|\Theta(10)|) \simeq 0.2$ where
$\Theta_{\PM}(10) \simeq -\Theta_{\PP}(10) \simeq 2.5$. 
Retardation causes the
van-der-Waals potential to decay as a function of $z$ more rapidly than
predicted by Eq.~\reff{eq:vdW}. This additionally reduces the
contribution of $\Phi_{0,{\rm vdW}}$ compared with $\Phi_C$. 
In the analysis of the experimental data in Sec.~\ref{sec:res} we shall
reconsider the Hamaker constant $A$ for the specific
system we are dealing with.

In order to achieve the accuracy of the temperature control 
needed for our measurements we have 
designed a cell-holder rendering a temperature stability of $\pm 5\,$mK. 
This has been accomplished by using a flow
thermostat coupled to the cell holder with a temperature of
$(30.50\pm 0.01)^\circ$C  functioning as a heat sink and shield
against temperature fluctuations of the environment. In order to fine tune
the temperature the cell was placed on a transparent ITO
(indium-tin-oxide) coated glass plate for a homogeneous heating of
the sample from below. The voltage applied to the ITO coating was
controlled via an Eurotherm  proportional-integral-derivative (PID) 
controller for approaching the demixing temperature. 
The controller feedback provides a
temperature stability of $\pm 2\,$mK at the position of the
Pt$100$-sensor used for temperature measurements. However, since
the probe particle is displaced from the sensor by a few
millimeter, some additional fluctuations have to be considered.
From the reproducibility of the potential measurements and from the
relative temperature fluctuations of two independent Pt100
sensors placed on either side of the cell we inferred a $\pm 5\,$mK
stability of the temperature at the actual position of the
measurement. The highly temperature sensitive measurements were
affected neither by the illuminating nor by the tweezing lasers
due to moderate laser powers and due 
to low absorption by the probe particle
and by the surface. 
Although all measurements were carried out 
upon approaching the demixing temperature, the light which is
increasingly scattered in the bulk background 
by the correlated fluctuations of the binary liquid mixture
exposed to the evanescent field turned out to be negligible compared to the
light scattered directly by the colloid. 
The effects of the onset of critical opalescence are significantly reduced by
the fact that the illuminating optical field rapidly vanishes upon increasing
the distance from the substrate and due to the still relatively small values of
the correlation length.

\subsection{The binary liquid mixture and boundary conditions}
\label{sec:exp_bm}

\begin{figure*}
\centering\includegraphics[width=1.3\columnwidth]{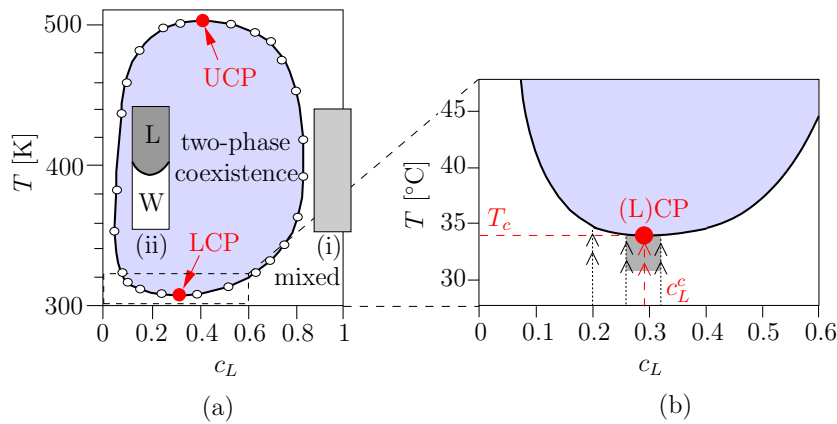}
\caption{Bulk phase diagram of the binary liquid mixture of water and
  2,6-lutidine (dimethylpyridine ${\rm C}_7{\rm H}_9{\rm
  N}$)~\cite{WL-phdiag,beysens1} at constant volume. 
  The relevant thermodynamic variables are the temperature $T$ and
  the mass fraction $c_L$ of lutidine in the mixture. Open symbols in panel
  (a) refer to actual experimental data~\cite{WL-phdiag}. 
  The schematic side view
  of a vertical sealed cell filled with the binary liquid mixture is shown by
  the insets (i) and (ii) of panel (a) for thermodynamic states outside and
  inside of the coexistence loop, respectively. 
  In (ii) W and L indicate the water-
  and the lutidine-rich phase, respectively. The mixture separates into the
  lutidine-rich and the water-rich phase within
  the two-phase coexistence area encircled by the solid first-order transition
  line. 
  At the lowest and highest points (LCP and UCP, respectively) of this line the
  demixing transition is continuous. The detailed 
  phase diagram in the vicinity of
  the lower critical point is shown in panel (b), together
  with the typical thermodynamic paths (dashed vertical lines) experimentally
  investigated here within the gray region as well as for $c_L=0.2$.
}
\label{fig:WL-PD}
\end{figure*}

For providing the critical fluctuations we have chosen 
the binary liquid mixture 
of water and $2,6$-lutidine near its demixing phase transition. 
The bulk phase diagram of such a binary liquid mixture prepared at room
temperature, ambient pressure, and sealed in a
cell~\cite{WL-phdiag} (constant volume) 
is reported in Fig.~\ref{fig:WL-PD}(a). 
It is characterized by a one-phase region (disordered phase) 
in which the two components form a mixed solution and which surrounds
the closed loop of the two-phase 
region (ordered phase) in which these components segregate into a
water-rich and a lutidine-rich phase. 
The first-order transition line delimiting the two-phase coexistence region,
within which the two ordered phases form an interface,
ends in a lower critical demixing point (LCP, see Fig.~\ref{fig:WL-PD}(a)) 
at the lutidine mass fraction
$c^c_L\simeq 0.28$ and the critical temperature $T_c \simeq
307.15\,$K~\cite{beysens1}. The upper critical demixing point (UCP, see
Fig.~\ref{fig:WL-PD}(a)) is located at high temperatures and therefore it 
occurs within the liquid phase only at pressure values above ambient ones. 

The choice of this specific binary liquid mixture as the critical medium 
is motivated by the fact that its properties (bulk phase
diagram, refractive index, viscosity, etc.) 
are known rather well and are
documented in the literature, as this mixture 
has been extensively employed in the past
for the study of phase separation per se 
or as a solvent of colloidal dispersions. 
A clear experimental advantage is provided by the fact that the
water-lutidine mixture at ambient pressure 
has a lower critical point slightly above 
room temperature which can be conveniently accessed from the one-phase 
region by \emph{heating} the sample. 
Alternative binary liquid 
mixtures with a lower demixing point close to room temperature
are formed by water and, e.g., 2,5-lutidine ($T_c\simeq 13^\circ$),
2,4-lutidine ($T_c\simeq 23^\circ$C), 
triethylamine ($T_c \simeq 18^\circ$C),
and $n$-butoxyethanol ($T_c\simeq 49^\circ$C). 
%
The addition of a suitable amount of a third component to some of these binary
liquid mixtures allows one to shift the critical temperature of demixing
basically at will. 
For example, by adding 3-methylpyridine to
the presently used mixture of 2,6-lutine and water it is possible to gradually 
shrink the coexistence loop in Fig.~\ref{fig:WL-PD}(a), causing an upwards
shift of $T_c$ of as much as $30^\circ$C. 
Analogously, even though 3-methylpyridine is always miscible with 
normal water, it
exhibits a coexistence loop if mixed with heavy water, with a lower
critical point at $T_c\simeq 39^\circ$C. The addition of normal 
water to this demixing binary
liquid 
mixture of heavy water and 3-methylpyridine 
causes an upwards shift of $T_c$ until the coexistence loop disappears
at a double critical point with 
$T_c^{\rm dcp}=78^\circ$C~\cite{WL-phdiag}.
%
The variety of available substances and the tunability of the critical
temperature by suitable chemical additions
allow one to generate critical Casimir
forces for a conveniently wide range of temperatures.

In our experiment the mixture is prepared under normal conditions (room
temperature, ambient pressure) and then it is 
introduced into the sample cell which is
afterwards sealed with Teflon plugs in order to hinder the mixture from
evaporating.
Although we have no control on the resulting pressure $p$ of the mixture, 
the fact that the cell is not hermetically sealed and that
small air bubbles might be trapped within it should keep $p$ 
very close to its ambient value. 
Within the limited range of temperatures we shall explore in the experiment,
possible pressure variations are not expected to lead to
substantial modifications
of the phase diagram (e.g., shifts of the critical point) compared to the ones
at constant pressure or volume.  
%
%
%
The order parameter $\op$ 
for the demixing phase transition can be taken to be the
difference between the local concentration $c_L({\bf x})$ 
({\it mass fraction}) 
of lutidine in the
mixture and its spatially averaged value $c_L$. 
Accordingly, a surface which
preferentially adsorbs lutidine is referred to as realizing the 
$(+)$ boundary condition for the order parameter given that it favors
$\op>0$, whereas a surface which preferentially adsorbs water leads to the 
$(-)$ boundary condition. 

The experimental 
cell containing the binary liquid 
mixture and the colloid is made up of silica
glass. Depending on the chemical treatment of its internal surface, one can
change the adsorption properties of the substrate so that it exhibits a clear
preference for either one of the two components of the binary mixture. In
particular, treating the surface with NaOH leads to preferential adsorption of
water $(-)$, whereas a treatment with hexamethyldisilazane (HMDS) favors the
adsorption of lutidine $(+)$~\cite{footnote_det_exp}, as we have
experimentally verified by comparing the resulting contact angles for water
and lutidine on these substrates.
%
%
As colloids we used polystyrene particles of nominal diameter 
$2R=3.69\,\mu$m and $2.4\mu$m, 
the latter possessing a rather high nominal surface charge
density of 10$\mu$C/cm$^2$. 
Size 
polydispersity of these particles are $2\%$ and
$3\%$, respectively, corresponding to ca. $\pm 70\,$nm. 
(These nominal values are provided 
in the data-sheets of the company producing the batch of particles, see
Ref.~\cite{nature} for details.) 
The adsorption properties of polystyrene particles
in a water-lutidine mixture have been investigated in 
Refs.~\cite{gm-92,gkm-92},
with the result that highly charged ($\gtrsim 3.8\mu$C/cm$^2$) 
colloids preferentially adsorb water (highly polar) whereas lutidine is
preferred at lower surface charges. 
Even though we did not independently determine these
adsorption properties, the results of Refs.~\cite{gm-92,gkm-92} and the
corresponding nominal values of the surface charges of the colloids  employed
in our experiment suggest that  the polystyrene particles of diameter
$2R=3.69\,\mu$m  [$2R=2.4\,\mu$m]  have a clear preference for lutidine $(+)$
[water $(-)$].
A posteriori, these presumed preferential adsorptions are consistent with the
resulting sign of the critical Casimir force observed experimentally. 
Depending on the surface treatment of the cell and the choice of the colloid
one can realize easily all possible combinations of (particle, substrate)
boundary conditions (see Tab.~\ref{tab:BC}).

\begin{table}
\centering
 \begin{tabular}{| c || c | c || }
\hline
\multicolumn{1}{|c||}{(particle, substr.)} 
& \multicolumn{2}{c||}{colloid diam. $2R$:} \\
\hline
    substrate treat.: & \phantom{ai}3.69$\mu$m\phantom{ai} & \phantom{ai}2.4$\mu$m\phantom{ai} \\ \hline\hline
    HMDS & $\PP$  & $\MP$ \\ \hline
    NaOH & $\PM$  & $\MM$ \\
    \hline
  \end{tabular}
\caption{Experimental realization of all possible 
  (particle, substrate) symmetry-breaking boundary
  conditions, where $(+)$ indicates the preferential adsorption of lutidine
  and $(-)$ the preferential adsorption of water. The treatments of the
  substrate affect only its surface properties.}
\label{tab:BC}
\end{table}

For a given choice of the particle-substrate combination with its 
boundary conditions and for a given 
concentration $c_L$ of the mixture we have determined the
interaction potential $\Phi$ (see Eq.~\reff{eq:potgen}) between the colloid
and the substrate as described in the previous subsection, starting from a
temperature $T$ below the critical point in the one-phase region and 
then increasing it towards 
that of the demixing phase transition line at this value of $c_L$. 
It might happen that, as a result of leaching, 
the water-lutidine mixture slowly (i.e., within several 
days) alters the surface
properties of the colloidal particles we used in the experiment. 
In order to rule out  a possible
degradation of the colloid during the experiment, we
verified the reproducibility of the observed effects after each
data acquisition.

\section{Results} 
\label{sec:res}

In, c.f., 
Figs.~\ref{fig:potmm}--\ref{fig:potmp}
and~\ref{fig:offcrit} we report the experimentally obtained 
interaction potentials $\Phi$ as
functions of the particle-wall distance $z$ for various values of the
temperature $T$, both at critical (Figs.~\ref{fig:potmm}--\ref{fig:potmp}) 
and off-critical concentrations (Fig.~\ref{fig:offcrit}).
In all the cases presented, the gravitational and offset parts of the
potentials [see Eq.~\reff{eq:potgen}], which turn out to be {\it de facto} 
independent of
the temperature $T$,  have been subtracted in such a way that
the resulting potentials vanish for large values of $z$. However, those
potentials, for which the gravitational tail could 
not be sampled  (see, e.g.,
Figs.~\ref{fig:potpp} and~\ref{fig:offcrit}), cannot be
normalized like the others by this requirement. 

Depending on the concentration of the mixture, two qualitatively different
behaviors are observed, which are discussed in Sec.~\ref{sec:res:cc} for
$c_L = c_L^c$ and in Sec.~\ref{sec:res:ncc} for $c_L \neq c_L^c$.
However, in the next subsection we first discuss
the experimental results for the potentials measured far away from the
transition line and the comparison of them with theoretical predictions. 
This provides important insight into the effective background forces to
which the critical Casimir forces add upon approaching the critical
point. 

\subsection{Non-critical potentials} 
\label{sec:res:zp}

\begin{figure}
\centering\includegraphics[width=\columnwidth]{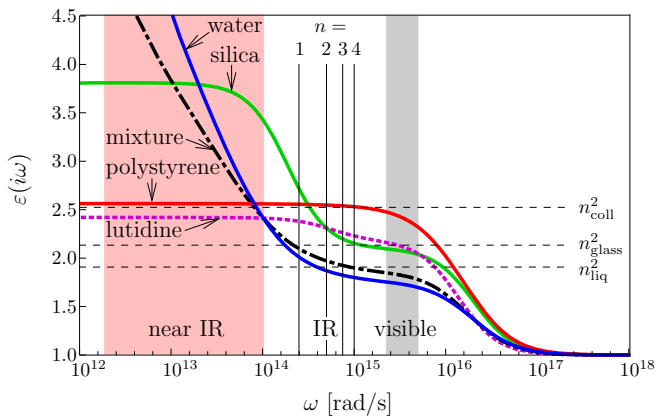}
\caption{Permittivities $\perm(i\omega)$ 
of the materials relevant for 
our experiment as functions of the frequency $\omega$ [rad/s] on a
logarithmic scale. As on the
left, from bottom to top, we report the curves corresponding to
pure lutidine, polystyrene ($\perm_{\rm coll}$), silica ($\perm_{\rm glass}$), 
a water-lutidine mixture with a volume
concentration $\phi_{\rm L} \simeq 0.25$  of lutidine ($\perm_{\rm liq}$), 
and pure water. 
These curves are based on the parameterizations and on the
material properties reported in Refs.~\cite{vdW,PLB-98}. The four vertical
lines indicate the frequencies $\{\omega_n\}_{n=1,\ldots 4}$ which within our
approximation enter into the
determination of the Hamaker constant in Eq.~\reff{eq:Ham}. (Note that 
$\mbox{Im\,}\perm(\omega)=0$ if, as in the present case, 
$\mbox{Re\,}\omega =0$~\cite{vdW}.)
}
\label{fig:epsilons}
\end{figure}

In all the cases reported here, sufficiently far from the transition line one
observes a potential which appears to consist only of the 
{\sl e}lectro{\sl s}tatic repulsion
between the colloid and the substrate and which can be 
fitted well by
\be
\Phi_0(z) = \kB T\, \rme^{-\kappa (z-z_{\rm es})}
\label{eq:el-rep}
\ee
where $\kappa^{-1}$ is the Debye screening length and $z_{\rm es}$ the value
of the distance $z$ at which $\Phi_0(z=z_{\rm es}) = \kB T$. ($z_{\rm es}$ is
expected to depend, inter alia, on the surface charge and on the radius 
of the colloid.)
For the potential in, c.f.,
Fig.~\ref{fig:potmm} which corresponds to $T_c-T = 300\,\mbox{mK}$, 
a fit of $\kappa$ yields $\kappa^{-1} = 12 \mbox{nm} \pm 3
\mbox{nm}$, which is compatible with the estimate 
$\kappa^{-1} \simeq 10\,$nm 
derived from the standard expression $\kappa =
\sqrt{e^2\sum_i\rho_i/(\perm_{\rm liq}(0)\kB T)}$ (see, e.g.,
Ref.~\cite{vdW}), where $e$ is the
elementary charge, $\perm_{\rm liq}(0)$ the static permittivity of the
mixture (see below), and $\rho_i$ the number density of ions 
assumed to be monovalent and estimated in
Ref.~\cite{gm-92} for the dissociation of a salt-free water-lutidine mixture.
%
%
Within the range of distances
$z$ sampled in our experiment there is no indication of the presence of an
attractive tail in $\Phi_0$, which on the other hand 
is generically expected to occur due to 
dispersion forces, described by a potential as given in Eq.~\reff{eq:vdW}.
In order to compare this experimental evidence with theoretical 
predictions, below we shall discuss the determination of the Hamaker
constant $A$
in Eq.~\reff{eq:vdW} on the basis of the dielectric properties of the
materials involved in the experiment. The relation between them is 
provided by (see, e.g., Ref.~\cite{vdW})
\be
\begin{split}
A(z) & \simeq \frac{3}{2} \kB T \sum_{n=0}^\infty \!' \,\,
\left.\frac{\perm_{\rm glass} -
  \perm_{\rm liq}}{\perm_{\rm glass} +
  \perm_{\rm liq}} \,
\frac{\perm_{\rm coll} -
  \perm_{\rm liq}}{\perm_{\rm coll} +
  \perm_{\rm liq}}\,
\right|_{i\omega_n}\!\!R_n(z) \\
& \equiv A_{n=0} + A_{n>0}
\end{split}
\label{eq:Ham}
\ee
where the permittivities $\perm$ of the various materials are
evaluated at the imaginary frequencies $i\omega_n$, with $\omega_n = 2\pi \kB
T n/\hbar = n\times 2.5 \times 10^{14}\,$rad/s at $T\simeq 300\,$K. 
(Note that the imaginary part of the 
complex permittivity $\perm(\omega)$ as a function of the complex frequency
$\omega$ vanishes on the imaginary axis $\mbox{Re\,} \omega=0$~\cite{vdW}.)
The factor $R_n(z)$ accounts for retardation and, neglecting the
fact that in the three different media light propagates with different
velocities (i.e., for $R_n$ assuming $\perm_{\rm coll} \simeq \perm_{\rm liq}
  \simeq \perm_{\rm coll}$)  it takes the form 
$R_n = (1+r_n)\rme^{-r_n}$.
The ratio $r_n \equiv 2 \tau_n(z)/ \omega_n^{-1}$  quantifies the relevance of
retardation: heuristically, a thermally fluctuating electric dipole within,
e.g., the glass generates an electric field which travels at least a distance
$z$ across the liquid, taking a minimal time $\tau_n(z) = z/(c/\sqrt{\perm_{\rm
    liq}(i\omega_n)})$, before inducing an electric dipole within the
colloid. Such an induced dipole, in turn, generates an electric field which
travels back to the original dipole and interacts with it. However, such an
interaction is reduced by the fact that the original dipole has a lifetime
$\omega_n^{-1}$ and might have decayed during the minimal time $2\tau_n(z)$
it takes the electric field to do the roundtrip~\cite{vdW}, which is the case
for $r_n \equiv 2 \tau_n(z)/ \omega_n^{-1} \gtrsim 1$.
The prime in Eq.~\reff{eq:Ham} indicates that the contribution of the static
permittivities $n=0$ is to be multiplied by $1/2$ (see, e.g.,
Ref.~\cite{vdW}), resulting in the term $A_{n=0}$.  
Within a first approximation, in Eq.~\reff{eq:Ham} 
the retardation factor $R_n$ 
does not affect those terms with $r_n\lesssim 1$, corresponding to
$R_n\simeq 1$ for them, while it exponentially 
suppresses those terms with $r_n\gg 1$. 
Accordingly, only the former contribute significantly to $A_{n>0}$ and
retardation is accounted for by summing in $A_{n>0}$ only the terms
corresponding to $\omega_n \lesssim (c/\sqrt{\perm(i\omega_n)})/(2 z)$, where
$\sqrt{\perm(i\omega_n)} \simeq n_{\rm liq}$ (see, c.f., dash-dotted line in
Fig.~\ref{fig:epsilons} for $n=1,\ldots,4$) 
is the refractive index of the liquid. 
The expression in Eq.~\reff{eq:Ham} is valid generically for a
film and only in the non-retarded
regime $R_n\simeq 1$ for the sphere-plate geometry. However, even for the
latter geometry an estimate of the order of magnitude of the effects of
retardation can be inferred simply by restricting the sum in Eq.~\reff{eq:Ham} 
to the values of $n$ corresponding to $r_n\lesssim 1$, so that $R_n\simeq 1$.

The parameters of our experiment, i.e., 
$z \gtrsim 0.1\,\mu$m and 
$n_{\rm liq} \simeq 1.38$ yield $\omega_n \lesssim 1.1
\times 10^{15}$ Hz, i.e., $n \le 4$ at $T\simeq 300\,$K, with 
$\omega_{n=1,\ldots,4}$ in the infrared (IR) spectrum. 
The contribution $A_{n=0}$  
of the zero-frequency mode is actually subject to
screening by the salt in the liquid solution, characterized by the
Debye screening length $\kappa^{-1}$ which also controls the exponential 
decay of the electrostatic contribution to $\Phi_0$. This means that $A_{n=0}$
is not a constant but acquires a $z$ dependence. This is 
accounted for by a multiplicative correction factor 
$R_0(z) = (1 + 2 \kappa z) \rme^{-2\kappa z}$ 
multiplying $A_{n=0}$ as given by Eq.~\reff{eq:Ham}; 
we note that $R_0(0)=1 \ge R_0(z)$.
A parametric representation of the permittivity $\perm_{\rm
  W}(i\omega)$  of pure water can
be found in Table~L2.1 in Ref.~\cite{vdW}. 
In the part of the spectrum over which the sum in 
Eq.~\reff{eq:Ham} runs, for polystyrene the permittivity  $\perm_{\rm
  coll}(i\omega)$ is actually almost constant $\perm_{\rm coll} =
n_{\rm coll}^2 \simeq 2.53$ 
(see, e.g., Table~L2.3 in Ref.~\cite{vdW} for a parameterization). 
For lutidine and
silica, the parameters which characterize
the corresponding permittivities $\perm_{\rm L}(i\omega)$ and 
$\perm_{\rm glass}(i\omega)$ are
summarized in Table~1 of Ref.~\cite{PLB-98}. In order to calculate the
dielectric permittivity $\perm_{\rm liq}(i\omega)$ of the homogeneous 
water-lutidine mixture on the basis of $\perm_{\rm W}$, $\perm_{\rm L}$, and
the lutidine volume fraction $\phi_{\rm L} \simeq 0.25$,
one can use the Clausius-Mossotti relation, as explained in
Ref.~\cite{PLB-98}. The resulting permittivities are reported in
  Fig.~\ref{fig:epsilons}.
%
%
With these elements at hand and within the approximation discussed above 
one can calculate an upper bound to the value of $A_{n>0}$ which accounts for
retardation, finding
\be
A_{n>0}(z\gtrsim 0.1\,\mu\mbox{m}) \lesssim 0.06\, \kB T \simeq  0.025 \times
10^{-20} {\rm J}.
\label{eq:Anz}
\ee
In addition, from Eq.~\reff{eq:Ham} and from the material parameters one 
finds
\be
A_{n=0} \simeq 0.46\, \kB T 
\ee
resulting in a screened contribution $A_{n=0}^{\rm (scr)}(z) = A_{n=0}R_0(z)$,
with $\kappa^{-1} \simeq 12$nm, which is actually negligible compared to the
electrostatic potential $\Phi_0$ [see Eq.~\reff{eq:el-rep}]: $A_{n=0}^{\rm
  (scr)}(z)/\Phi_0(z) \lesssim A_{n=0} \rme^{-\kappa z_{\rm es}} \lesssim
6\times 10^{-4}$ (c.f., Tab.~\ref{tab:zp} for typical values of $\kappa$
and $z_{\rm es}$). Taking into account Eq.~\reff{eq:vdW}, the contribution of
$A_{n=0}^{\rm (scr)}$ to $\Phi_{0,{\rm vdW}}(z)$ becomes comparable to the
electrostatic one only for $z/R \lesssim 10^{-4}$, which is well below the
range $z/R \gtrsim 0.1$ actually investigated in our experiment.
On the other hand, 
for $z\simeq 0.1\,\mu$m, $A_{n=0}^{\rm (scr)}\simeq 0.004\,\kB T$ and due to
its exponential decay with $z$, we expect this contribution to be negligible
compared to the value $A_{n>0}(z\gtrsim 0.1\,\mu\mbox{m})$ in
Eq.~\reff{eq:Anz}.
In turn, this value is considerably
smaller than the typical one $A\simeq 10^{-20}$J 
we have used in Sec.~\ref{subsec:IP} in order
to compare dispersion forces with the critical Casimir potential. Accordingly,
the conclusion drawn there that the latter typically dominates the former is
confirmed and reinforced by the estimate for $A$ given here.
The values just determined for $A_{n=0}$ and $A_{n>0}$ are meant to be
estimates of their orders of magnitude given that a
detailed calculation which properly accounts for retardation (following, e.g.,
Ref.~\cite{PLB-98}) and for possible inhomogeneities in the media, especially
within a
binary liquid 
mixture, goes beyond the present scope of a qualitative comparison of
theoretical predictions for the background forces with
the actual experimental data. In this respect, a detailed determination of the 
permittivities of the specific materials used in our experiment would be 
crucial for an actual quantitative comparison of this contribution to $\Phi$
with the experimental data. 
With all these
limitations, the theoretical calculation discussed above yields $A(z\gtrsim
0.1\,\mu\mbox{m})\simeq A_{n>0}(z\gtrsim 0.1\,\mu\mbox{m}) \lesssim 0.06\, \kB
T$.
If one insists on fitting the experimental 
data for the background potential $\Phi_0(z)$ by
including the contribution of
the dispersion forces as given by Eq.~\reff{eq:vdW} in addition to a possible
overall shift $\Delta\Phi_{\rm offset}$, 
one finds values for the Hamaker constant $A$ which vary as
function of the range of values of $z$ which the considered 
data set refers to.  %
This might be due to the fact that the statistical error affecting the
data increases at larger distances or due to an incomplete subtraction of the
gravitational contribution, which might bias the result. 
In particular,  in the range $0\le z \le
0.3\,\mu$m we focus on data for the potentials which have been measured
experimentally for the largest temperature deviation from the critical point
and which are smaller than $6 \,\kB T$. The choice of this latter 
value results from
a compromise between avoiding the increasing
statistical uncertainty due to the poor sampling of the 
sharply increasing potential
and having a sufficiently large number of data points left at short distances,
where $\Phi_0(z)$ is not negligible.
The resulting parameter values for the four experimentally measured
potentials are reported in Tab.~\ref{tab:zp}. 
%
%
\begin{table*}
\centering
\begin{tabular}{|c|c|c||c|c|c|c|}
\hline
BC & Fig. & $T_c-T[K]$ &$A[\kB T]$ & $\Delta\Phi_{\rm offset}[\kB T]$ &
$\kappa^{-1}[{\rm nm}]$ & $z_{\rm es}\,[\mbox{nm}]$\\
\hline
\hline
$\MM$ &\ref{fig:potmm}& 0.30 & $0.2\pm0.1$ & $0.14\pm0.08$ & $10.5\pm0.5$ &
$113\pm 1$\\
$\PM$ &\ref{fig:potpm}& 0.90 & $0.2\pm0.1$ & $0.2\pm0.1$ & $17\pm1$ &
$90\pm 3$\\
$\PP$ &\ref{fig:potpp}& 0.20 & $0.05\pm0.03$ & $0.06\pm0.04$ & $15.9\pm0.5$ &
$85\pm 1$\\
$\MP$ &\ref{fig:potmp}& 0.31 & $0.0\pm0.2$ & $0.0\pm0.1$ & $13\pm1$ &
$ 153\pm 2$\\
\hline
\end{tabular}
\caption{Fit parameters for 
the non-critical potentials $\Phi_{\rm non-cr}$ 
for four boundary conditions
  and with the gravitational part subtracted, 
  $\Phi_{\rm non-cr}=\Phi_{0,{\rm vdW}}(z) + \Delta\Phi_{\rm
  offset} + \kB T\, \rme^{-\kappa (z-z_{\rm es})}$, 
where $\Phi_{0,{\rm vdW}}(z)$ is
  given by Eq.~\reff{eq:vdW}. The
  values reported here correspond to 95\% confidence intervals for 
  the parameters. 
\label{tab:zp}}
\end{table*}
%
%
The resulting values of $A$ are compatible with a rather small
Hamaker constant, in qualitative agreement with the previous theoretical
analysis. The combined estimate of the screening length is somewhat larger
than anticipated from the analysis of one of the potentials [see after
Eq.~\reff{eq:el-rep}] and results in $\kappa^{-1} = (14\pm4)\,$nm, again in
agreement with independently available experimental data~\cite{gm-92}.
In order to highlight the presence of dispersion forces in this system,
here masked by the strong electrostatic repulsion, one would have to
increase the salt concentration of the solvent 
in order to reduce significantly the screening length $\kappa^{-1}$ which then
provides access to smaller particle-substrate distances. However, we emphasize
that  a detailed and quantitative study of these background forces is not
necessary in order to identify the contribution of critical Casimir forces to 
the total potential and it is therefore beyond the scope of the present 
investigation. 

\subsection{Critical composition} 
\label{sec:res:cc}

\subsubsection{Experimental results}

\begin{figure}
\centering\includegraphics[width=\columnwidth]{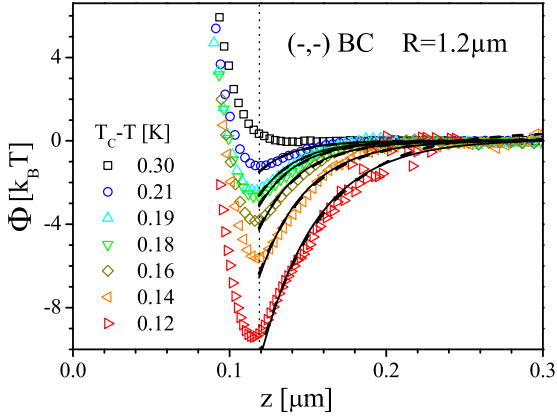}
\caption{%
Effective interaction potential $\Phi(z)$ between a wall and a spherical 
particle of radius $1.2\,\mu$m immersed in a
water-lutidine mixture at the critical concentration $c^c_L$, as a function of
the distance $z$ from the wall and for various values of the temperature 
in the one-phase region ($T<T_c$)~\cite{nature}. 
The gravitational and the offset contribution to the potential
[see Eq.~\protect{\reff{eq:potgen}}] have been subtracted. 
The set of solid and dashed lines, which are barely distinguishable on this
scale, correspond to the theoretical predictions (see the main text for
details). 
The potentials reported here refer to the $\MM$ boundary conditions
(other cases are reported in Figs.~\protect{\ref{fig:potpm}}
and~\protect{\ref{fig:potpp}}) and show that an increasingly attractive
force contributes to the total potential upon approaching the critical
temperature, i.e., upon decreasing $\Delta T = T_c - T$. Here $T_c$ is the
nominal value of the critical temperature, corresponding to the anomaly in the
background scattering, which signals the onset of critical opalescence in the
sample.  
Only the data to the right of the vertical dotted line are
considered for the comparison with the theoretical predictions (see the 
main text). 
}
\label{fig:potmm}
\end{figure}

\begin{figure}
\centering\includegraphics[width=\columnwidth]{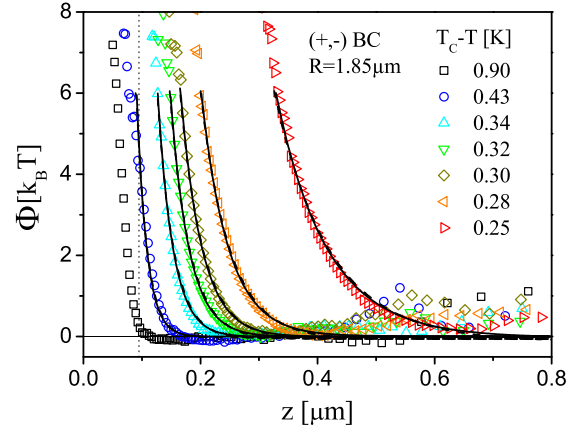}
\caption{%
Interaction potential $\Phi(z)$ as in Fig.~\protect{\ref{fig:potmm}} for
the $\PM$ boundary conditions and $R\simeq 1.85\,\mu$m~\cite{nature}. 
An increasingly repulsive
force contributes to the total potential upon approaching the critical
temperature. 
The set of solid and dashed lines, which are barely distinguishable on this
scale, correspond to the theoretical predictions (see the main text for
details).  Only the data to the right of the vertical dotted line are
considered for the comparison with the theoretical predictions (see the 
main text). 
}
\label{fig:potpm}
\end{figure}

\begin{figure}
\centering\includegraphics[width=\columnwidth]{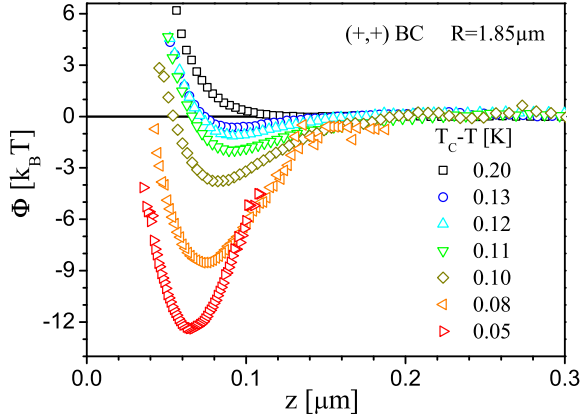}
\caption{%
Interaction potential $\Phi(z)$ as in Figs.~\protect{\ref{fig:potmm}}
and~\protect{\ref{fig:potpm}},  for the $\PP$ boundary
conditions~\cite{nature}.
As for $\MM$ boundary conditions (see
Fig.~\protect{\ref{fig:potmm}}), 
upon approaching the critical temperature
an increasingly attractive force contributes
to the total potential.  
}
\label{fig:potpp}
\end{figure}

\begin{figure}
\centering\includegraphics[width=\columnwidth]{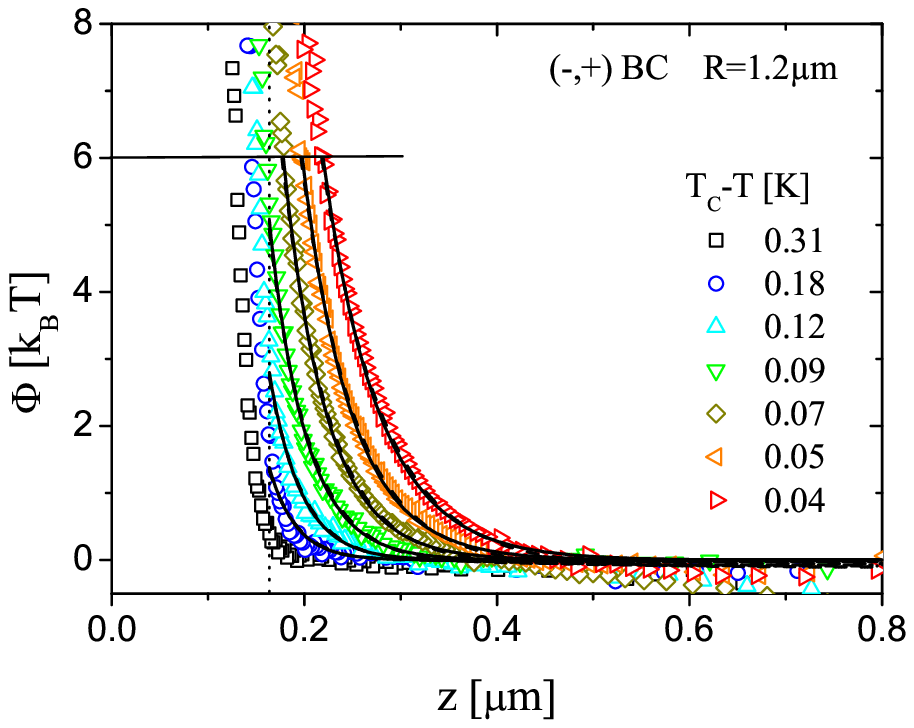}
\caption{%
Interaction potential $\Phi(z)$ as in
Figs.~\protect{\ref{fig:potmm}}, \protect{\ref{fig:potpm}}, 
and~\protect{\ref{fig:potpp}} for the $\MP$ boundary conditions. 
As in the case of the $\PM$ boundary conditions (see
Fig.~\protect{\ref{fig:potpm}}) an increasingly repulsive force contributes
to the total potential upon approaching the critical temperature. 
The set of solid and dashed lines, which are barely distinguishable on this
scale, correspond to the theoretical predictions (see the main text for
details).   Only the data to the right of the vertical dotted line are
considered for the comparison with the theoretical predictions (see the 
main text). 
}
\label{fig:potmp}
\end{figure}

For the binary liquid 
mixture at the critical composition we have estimated 
(after data acquisition) the
critical temperature $T_c$ as the temperature at which 
anomalies in the background light 
scattered by the mixture 
in the absence of the colloid and 
due to critical opalescence are
observed and visual inspection of the sample displays an incipient phase
separation.
The value determined this way has to be
understood as an estimate of the actual value of the critical temperature
of the water-lutidine mixture and it is used for the calibration of the
temperature scale, which is shifted in order to set $T_c$ to the nominal 
value $T_c = 307.15\,$K reported in the literature (see, e.g.,
Ref.~\cite{beysens1}). Note, however, that depending on the different level of
purity of the mixture, published experimental values of $T_c$ 
are spread over the range $306.54\div 307.26$\,K (see, e.g., the summary in
Ref.~\cite{mb06}). 
Due to the difficulties in determining 
the absolute value of the critical temperature, 
with our experimental setup  only temperature
differences are reliably determined and the actual
critical temperature of the mixture might differ slightly from the nominal
value $T_c$.  We shall account for this fact in our comparison 
with the theoretical predictions. 

Close to $T_c$ 
critical opalescence is expected to occur. It is indeed ultimately 
observed upon heating the mixture towards
the critical temperature, leading to an increase in the background light
scattering due to the correlated fluctuations in the mixture. Even though this
might interfere with the determination of the interaction potential $\Phi$ via
TIRM, within the range of temperatures we have explored at the critical
concentration, the enhancement in the background scattering is actually
negligible compared to the light scattered by the particle.
%
%

In Fig.~\ref{fig:potmm} we present the interaction potentials $\Phi$ as a
function of the distance $z$ for that choice of colloidal particle and surface
treatment which realizes the $\MM$ boundary condition (see
Tab.~\ref{tab:BC}). 
As discussed above, for $\Delta T \equiv T_c-T = 0.30$K, the potential consists
only of the electrostatic repulsion (see Eq.~\reff{eq:el-rep}). Upon
approaching the critical point an increasingly deep potential well gradually
develops, indicating that an increasingly strong {\it attractive} force is
acting on the particle. At the smallest $\Delta T$ we have investigated, i.e.,
$\Delta T = 0.12$K, the resulting potential well is so deep 
that the particle hardly escapes from it. 
In view of the small temperature 
variation of ca.~180\,mK, 
the change of ca.~$10 \kB T$ in the resulting potential is remarkable. 
This very sensitive dependence on $T$ is a clear indication that in the
present case 
critical Casimir forces are at work.
In the case of Fig.~\ref{fig:potmm} the maximum attractive force acting on the
particle is about 600\,fN.

According to the theoretical predictions, one expects the critical Casimir
force to be {\it repulsive} for asymmetric boundary conditions $\PM$ or
$\MP$. In our experiment we can easily realize the $\PM$ BC by changing the
colloidal particle surface from preferentially adsorbing water $(-)$ to
preferentially adsorbing lutidine $(+)$, without any additional 
surface treatment of the cell (see Tab.~\ref{tab:BC}).
The interaction potentials $\Phi$ for this case are reported in
Fig.~\ref{fig:potpm}. As for the $\MM$ boundary condition, sufficiently
far from the critical point (i.e., $\Delta T = T_c-T = 0.90\,$K), 
the potential consists only of the electrostatic repulsion contribution [see
Eq.~\reff{eq:el-rep}]. 
However, upon approaching the critical point the
repulsive part of the potential curves shifts towards larger values of the
distance $z$, indicating that an additional {\it repulsive} force is acting on
the colloid. 
It is possible to make this force attractive again by treating the surface of
the cell so that its preferential adsorption changes from water $(-)$ to
lutidine $(+)$, so that then $\PP$ boundary condition is realized. As
Fig.~\ref{fig:potpp} shows, the resulting potentials show indeed an
attractive part the qualitative features of which resemble those of the case
with the $\MM$
boundary condition, reported in Fig.~\ref{fig:potmm}. Note that the depth of
the potential in Fig.~\ref{fig:potpp} corresponding to $\Delta T = 0.05$ is so
large that the gravitational part (which has been subtracted) cannot be
sampled by the particle and therefore the position of this potential
curve along the vertical axis cannot be fixed. 
If, with the same  $(+)$ surface of the cell, one changes again the colloidal
particle from preferentially adsorbing lutidine $(+)$ to preferentially
adsorbing water $(-)$, we can experimentally realize the $\MP$ boundary
condition (see Tab.~\ref{tab:BC}) 
for which a repulsive critical Casimir force is
expected. The resulting potential is reported in Fig.~\ref{fig:potmp} and
shows the same qualitative features as the one in Fig.~\ref{fig:potpm}, with
an increasingly repulsive force which builds up upon approaching the critical
point.

\subsubsection{Comparison with theory}

\label{sss:ct}


The experimental data reported in the previous subsection can be compared
with the theoretical predictions presented in Sec.~\ref{sec:th:cc}, which
are expected to be valid for 
$\delta \equiv z/R \ll 1$ (Derjaguin approximation). 
In the experimental setting corresponding to Figs.~\ref{fig:potmm}
and~\ref{fig:potmp}, $R=1.2\mu$m whereas 
$z \lesssim 0.3 \mu$m and $z \lesssim 0.8\mu$m, respectively, so that $\delta
\lesssim 0.25$ and $\delta \lesssim 0.67$. In Figs.~\ref{fig:potpm}
and~\ref{fig:potpp} one has $R\simeq
1.85\mu$m with $z\lesssim 0.8$ and $z\lesssim 0.3$, respectively, corresponding
to  $\delta \lesssim 0.43$ and $\delta \lesssim 0.16$. 
Accordingly the Derjaguin
approximation is expected to provide a sufficiently accurate description of
the experimental data, 
possibly apart from those at larger values of $z$ in
Figs.~\ref{fig:potpm} and~\ref{fig:potmp}, 
the corresponding potential values of which are anyhow negligibly small.
In order to extract from the measured potential only the part which is due to
the critical Casimir force we focus on that range of distances $z$ for
which the electrostatic contribution $\Phi_0(z)$ 
(see Eqs.~\reff{eq:potgen} and~\reff{eq:el-rep}) as
measured far from the critical point (i.e., for $\Delta T = 0.30$K in
Fig.~\ref{fig:potmm} and $\Delta T = 0.90$K in Fig.~\ref{fig:potpm}) 
is actually negligible, using as a criterion $|\Phi_0(z)| \lesssim 0.5 \kB T$.
This latter choice also avoids additional complications due to possible
changes induced by critical fluctuations in the electrostatic contribution
$\Phi_0(z)$ upon approaching the critical point. 
Accordingly, for a quantitative comparison with the theoretical predictions
we consider only data corresponding to $z \ge 0.12\mu$m in
Fig.~\ref{fig:potmm} [$\MM$ BC], $z \ge 0.1\mu$m in
Fig.~\ref{fig:potpm} [$\PM$ BC], and $z \ge 0.16\mu$m in
Fig.~\ref{fig:potmp} [$\MP$ BC], excluding in each case 
the data set corresponding to the largest value of $\Delta T$, which has been
used to define $\Phi_0(z)$.
Unfortunately, the number of data points which
satisfy this condition in the case of Fig.~\ref{fig:potpp} is quite
limited for providing a basis for a reliable analysis; 
therefore we do not process these corresponding data. 

The strength of the critical Casimir force depends strongly on the deviation
$T_c-T$ from the critical point via the bulk correlation length
$\xi \sim |T_c-T|^{-\nu}$. %
Accordingly, even a small systematic error in the experimental determination
of $T$ and $T_c$ can result in sizeable discrepancies in the comparison 
between measured potentials and theoretical predictions.
Statistical variations and 
a possible drift of the temperature during the acquisition of the data, which
are kept within $5\,$mK by the temperature controller used in our experimental
apparatus~\cite{nature}, are similarly important.
In order to circumvent parts of these problems
we compare the experimental data, selected by the aforemention criterion (see
the vertical lines in Figs.~\ref{fig:potmm}, \ref{fig:potpm},
and~\ref{fig:potmp}), for a certain boundary condition and for the six 
temperatures 
$T_i = T_c - \Delta T_i$, $i=1$,\ldots, 6, closest to $T_c$ with 
the theoretical prediction for 
$\Phi_C(z;\xi)$
provided by Eqs.~\reff{eq:PhiDgen} and~\reff{eq:scalingfunc}. For each $T_i$
the values $\xi_i$ of the correlation length and of a possible
residual offset $\Phi_{{\rm offset},i}$ are determined in such a way as to
optimize the agreement between 
$\Phi_C(z;\xi_i) + \Phi_{{\rm offset},i}$ and the corresponding experimental
data set.
A drift of the temperature during the acquisition of the data might
affect the value of the correlation length $\xi_i$ resulting from this
procedure. As we shall see below, even if present, this possible drift does
not strongly affect the final estimate for the correlation length amplitude
$\xi_0$ in Eq.~\reff{eq:xidiv}, the uncertainty of which is
dominated by the systematic uncertainty of the theoretical
predictions stemming from finite-size extrapolations of the Monte Carlo data. 

The data set $(T_i,\xi_i)$ is then fitted with the theoretically expected
power-law behavior given by Eq.~\reff{eq:xidiv}. This is carried out by fixing 
the universal critical exponent $\nu\simeq
0.630$ to its best known 
theoretical value while determining the non-universal amplitude
$\xi^{\rm (fit)}_0$ and the value $T_c^{\rm (fit)}$ of the critical
temperature from the data set $(T_i,\xi_i)$:
\be
\xi_i = 
\xi^{\rm (fit)}_0 \left(1 -  \frac{T_i}{T_c^{\rm (fit)}}\right)^{-0.63} \,.
\label{eq:fitxi}
\ee
Here we assume that the temperatures 
$T_i$ are sufficiently close to $T_c$ so that $\xi$ is
described correctly by its \emph{leading} power-law behavior.
The resulting value of $\xi^{\rm (fit)}_0$ can then be compared with the
available independent experimental estimates reported in Tab.~\ref{tab:xi0},
providing a check of the consistency of the experimental data with the
theoretical predictions.
\begin{table}
\begin{tabular}{|ll||c|c|}
\hline
\multicolumn{2}{|c||}{$\xi^{\rm (exp)}_0\,$[\AA]} & Ref. (year) & Method \\
\hline\hline
$2.0\phantom{0}$&$\pm 0.2$ & \cite{gcsp72} (1972) & static LS\\
$2.92$&$\pm0.19$ & \cite{gcsp72} (1972) & dynamic LS, linewidth \\
$2.7\phantom{0}$&$\pm 0.2$ & \cite{jbww87} (1987) & static LS \\
$2.3\phantom{0}$ && \cite{elv93} (1993) & specific heat \\
$2.1\phantom{0}$ && \cite{slsl97} (1997) & critical adsorption\\
$1.98$&$\pm0.04$ & \cite{mb06} (2006) & specific heat\\
\hline
\end{tabular}
\caption{Experimental estimates of the non-universal 
correlation length amplitude $\xi_0$ for the water-lutidine mixture at the
critical concentration. In light scattering (LS) experiments 
the bulk correlation length
$\xi(T)$ is determined by measuring  the wave-vector (static) or frequency
(linewidth, dynamics) dependence of the scattered intensity. 
A fit of $\xi(T)$ to the expected algebraic behavior
[Eq.~\protect{\reff{eq:xidiv}}]
yields the value $\xi_0$.  Alternatively, $\xi_0$ can be obtained on the basis
of the measured value of the non-universal 
amplitude $A^+$ which characterizes the
divergence of the specific heat at constant pressure $C_p(\tau\rightarrow 0^+)
\simeq (A^+/\alpha)\tau^{-\alpha}$ and the theoretical~\protect{\cite{mb06}} or
experimental~\protect{\cite{elv93}} value of the universal amplitude ratio 
$Q^+ = A^+\xi_0^3\rho/(M\kB)$, where $M$ is the molar mass of the mixture
and $\rho$ its mass density at the critical point~\protect{\cite{Priv,PV}}. A
careful theoretical analysis of experimental data for the critical adsorption
profiles also leads to an estimate for
$\xi_0$~\protect{\cite{slsl97}}. Comparing the experimentally 
measured potentials to the theoretical predictions for the critical Casimir
contribution we obtain the estimate
reported in, c.f., Eq.~\reff{eq:xi0fit}.
} 
\label{tab:xi0}
\end{table}
%
%
In spite of the scattering of the available 
experimental data, which might be due to
different conditions of the mixture (such as contaminations or slightly
different concentrations) or to different systematic uncertainties of the
various approaches, all the estimates are within the range
\be
\xi^{\rm (exp)}_0 = 2.3\pm 0.4\, \mbox{\AA},
\label{eq:xi0exp}
\ee
estimated via a least-square fit of the
data~\cite{mb06,gcsp72,jbww87,elv93,slsl97} 
in Tab.~\ref{tab:xi0}.
(The experimental value quoted in Ref.~\cite{nature} refers to the estimate
of Ref.~\cite{gcsp72}.)
The limited set of temperatures which have been investigated experimentally
(apart from one far away from $T_c$ and used for fixing the background
potential, six different values for each set of boundary conditions) 
does not allow us to determine simultaneously and 
reliably the exponent $\nu$ from the experimental data.
%

%
In comparing the experimental data with the 
theoretical predictions we have to take into account the uncertainties which
affect both of them. 
As discussed in Sec.~\ref{sec:th:cc} the currently available theoretical
predictions within the Derjaguin approximation are affected by a 20\%
{\it systematic} uncertainty, clearly visible in Fig.~\ref{fig:DerjPot}, 
for the amplitude of the 
scaling function of the Casimir potential $\Phi_C$. 

As far as the experimental data are concerned, the
{\it systematic}  uncertainties --- 
which are the ones most relevant for the comparison --- concern
(i) the particle-wall distance $z$,
which can be determined by the hydrodynamic method up to 
$\Delta z =\pm 30$nm (see
Sec.~\ref{sec:exp_TIRM} and Ref.~\cite{footnote1}) 
and (ii) the absolute temperature scale $T$ and, in
particular, the value of the critical temperature $T_c$. As described above, in
order to cope with the uncertainty in $T$
we opted for an indirect determination of the associated correlation length
from the best fit of the experimental data with theoretical predictions. 
%

In addition to these systematic uncertainties, there are {\it statistical
errors} associated with the way the potential $\Phi$ is determined via TIRM. 
The number $N(n_{\rm sc})$ of counts during 
the sampling time $t_{\rm samp}$, registered in
each bin of size $\Delta n_{\rm sc}$, centered around $n_{\rm sc}$ and forming 
the intensity histogram $p_{\rm sc}(n_{\rm sc}) = 
N(n_{\rm sc})/(N_{\rm tot}\Delta n_{\rm sc})$ 
reported in Fig.~\ref{fig:data_analy}(b), 
is subject to statistical fluctuations $\Delta N(n_{\rm sc})$ 
which affect the estimate of $p_{\rm sc}(n_{\rm sc})$, $p_z(z)$, and
therefore of the potential $\Phi$; $N_{\rm tot} = f_{\rm samp}t_{\rm
  samp}$ is the total number of counts in the time series of $n_{\rm sc}(t)$ of
duration $t_{\rm samp}$, from which the histogram of $p_{\rm sc}(n_{\rm sc})$
has been constructed.
One expects that 
these statistical fluctuations are relatively more important for
those bins which are less populated, i.e., for smaller $N(n_{\rm sc})$. 
In terms of
the distance $z$ of the colloid from the wall, they correspond to values which
are less frequently sampled during the Brownian motion of the particle under
the influence of the
potential $\Phi$, i.e., to larger values of the potential.
This can be seen directly from the potentials reported in
Figs.~\ref{fig:potmm}, \ref{fig:potpm}, \ref{fig:potpp}, \ref{fig:potmp}, 
and, c.f., \ref{fig:offcrit}, 
in which the experimental data are more scattered very
close to the wall and far from it, whereas the sampling of the potential
$\Phi$ is particularly accurate around its minimum. 
In order to evaluate the statistical uncertainty associated 
with each data point
of the potential, ideally 
one should construct the histogram of  $p_{\rm sc}(n_{\rm sc})$ based on
several different realizations of the time series $n_{\rm sc}(t)$ and then
analyze the statistical properties of this ensemble of plots. 
Alternatively, one might evaluate the autocorrelation time $t_{\rm corr}$ 
of $n_{\rm sc}(t)$ [e.g., from a detailed study of the autocorrelation function
$C(\delta t)$, see the text before Eq.~\reff{eq:Dapp}]. 
Assuming that the number $N_{\rm in}(n_{\rm sc})$ of statistically
independent counts in a bin of the histogram is given by 
$N_{\rm in}(n_{\rm sc}) = N(n_{\rm sc})/(f_{\rm samp} t_{\rm corr})$ 
and assuming 
that the statistics of
the counts is Poissonian, the associated relative statistical
fluctuation is related to the number of counts by 
$\Delta N_{\rm in}(n_{\rm sc}) = \sqrt{N_{\rm in}(n_{\rm sc})}$ 
and induces a statistical
uncertainty $\Delta\Phi(z(n_{\rm sc}))/(\kB T) = \Delta N_{\rm in}(n_{\rm
  sc})/N_{\rm in}(n_{\rm sc}) = N^{-1/2}_{\rm in}(n_{\rm sc})$ for the value of
the potential $\Phi$ at the position $z(n_{\rm sc})$ corresponding to the
scattered number of photons $n_{\rm sc}$.
However, in the comparison between the experimental data and the theoretical
predictions, the statistical error in the former is expected to be negligible
compared to the systematic uncertainty in the latter and therefore we do not
proceed to a detailed evaluation of the statistical error associated 
with the
data points reported in Figs.~\ref{fig:potmm}, \ref{fig:potpm},
\ref{fig:potpp}, \ref{fig:potmp} , and, c.f., \ref{fig:offcrit}. 
Actually, a good estimate of the magnitude of the statistical error 
can be inferred from the scatter
of the experimental data points relative to a smooth curve interpolating
each potential. 

\begin{figure}
\centering\includegraphics[width=\columnwidth]{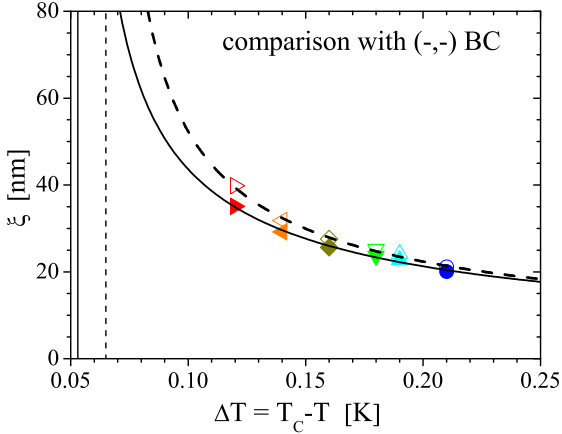}
\caption{%
Correlation length $\xi$ as a function of the temperature 
deviation $\Delta T = T_c-T$
from the experimentally located critical temperature $T_c$ for the
water-lutidine mixture at the critical concentration.
The sets of data points are obtained by optimizing the agreement
between the experimental data from Fig.~\ref{fig:potmm} 
and the corresponding theoretical predictions for the
Casimir potential for the $\MM$ boundary conditions (Fig.~\ref{fig:DerjPot}, 
see the main text for details). The upper (lower) 
set of points has
been determined by using as the scaling function $\Sf_{\MM} = \Sf_{\PP}$ 
of the Casimir potential the one
reported by the dashed (solid) line in Fig.~\protect{\ref{fig:DerjPot}}.
The dashed and solid curves are the best fit of the corresponding data sets 
based on Eq.~\protect{\reff{eq:fitxi}} with the corresponding values 
$\Delta T_c^{\rm (fit)} = T_c -
T_c^{\rm (fit)}$ indicated by the vertical dashed and solid lines,
respectively, i.e., $\xi$ diverges at $T_c^{\rm (fit)}$ corresponding to 
$\Delta T(T=T_c^{\rm (fit)}) = T_c - T_c^{\rm (fit)} =\Delta T_c^{\rm (fit)}$.
}
\label{fig:fitximm}
\end{figure}

In order to discard those data 
which are affected by large statistical fluctuations, we consider for the
comparison with the theoretical prediction only data fulfilling  
$\Phi(z) < 6 \kB T$ and  $z\le z_{\rm max}$ where 
$z_{\rm
max} = 0.3\,\mu$m
for 
Fig.~\ref{fig:potmm},  
\begin{center}
\begin{tabular}{|c|c|c|c|c|c|c|}
\hline
$z_{\rm max}$ [$\mu$m] &  0.3 &  0.35 &  0.35 & 0.4 & 0.5 & 0.8 \\
\hline
$\Delta T$[K] &   0.43 & 0.34 & 0.32 & 0.30 & 0.28 & 0.25 \\
\hline
\end{tabular}
\end{center}
for Fig.~\ref{fig:potpm}, and 
\begin{center}
\begin{tabular}{|c|c|c|c|c|c|c|}
\hline
$z_{\rm max}$ [$\mu$m] &  0.4 &  0.4 &  0.45 & 0.5 & 0.55 & 0.6 \\
\hline
$\Delta T$[K] &   0.18 & 0.12 & 0.09 & 0.07 & 0.05 & 0.04 \\
\hline
\end{tabular}
\end{center}
for Fig.~\ref{fig:potmp}.
For the $\MM$ boundary conditions, in Fig.~\ref{fig:potmm} we report the
comparison between the theoretical prediction and those experimental data
which have been selected as explained above.
The solid
and dashed lines, barely distinguishable, 
correspond to the predictions given by 
Eq.~\reff{eq:PhiDgen} by using for $\Theta_{\MM} = \Theta_{\PP}$ 
the scaling functions described by the solid and dashed lines, respectively, in
Fig.~\ref{fig:DerjPot}. 
(For each temperature $\Delta T_i$ the experimental
potentials have been shifted vertically 
by the amount $-\Phi_{{\rm offest},i}$ determined previously as
the best fit parameter.)
The corresponding values $\xi_i$  of the
correlation length are reported in Fig.~\ref{fig:fitximm} together
with the resulting best fit based on Eq.~\reff{eq:fitxi},
%
%
which leads to the least-square estimates 
$\xi_0^{\rm (fit)} = 1.7\pm 0.1$\AA\ and
$\Delta T_c^{\rm (fit)} \equiv T_c - T_c^{\rm (fit)} = 52\pm 10$ mK 
for the solid curve and  
$\xi_0^{\rm (fit)} =  1.7 \pm 0.1$\AA\ and
$\Delta T_c^{\rm (fit)} = 65\pm 7$ mK
for the dashed one.
Taking into account the systematic uncertainty of the scaling functions for the
Casimir potential, we arrive at the combined estimate
$\Delta T_c^{\rm (fit)} = 60 \pm 15$ mK, i.e., the value of $T_c$
determined experimentally is actually higher than the value 
$T_c^{\rm (fit)}$ resulting from
the comparison with the theoretical predictions.
In addition, this comparison allows one to estimate the correlation length
$\xi$, for which no independent experimental estimate is presently
available. 
According to Fig.~\ref{fig:fitximm} one has $20 \mbox{nm} \lesssim \xi \lesssim
40 \mbox{nm}$, so that for the range
$0.12\mu\mbox{m}\le z \le 0.3\mu\mbox{m}$ of distances this translates into the
ranges  $3 \lesssim x = z/\xi \lesssim 15$ and   
$6 \lesssim u = (1-T/T^{\rm (fit)}_c)^{1/\nu}(z/\xi_0^{\rm
  (fit)})^{1/\nu} \lesssim 70$ of the scaling variables $x$ and $u$ (see
Fig.~\ref{fig:DerjPot}). 
In order to be able
to test prominent features of the theoretically predicted scaling
function such as the occurrence of a minimum for $u_{\rm min} \simeq 0.5$, 
one has to reach $\xi \gtrsim 180$ nm, i.e., one must get still 
closer to the critical
point ($\Delta T \lesssim 6\,$mK) than it was possible in the 
present experiment.

\begin{figure}
\centering\includegraphics[width=\columnwidth]{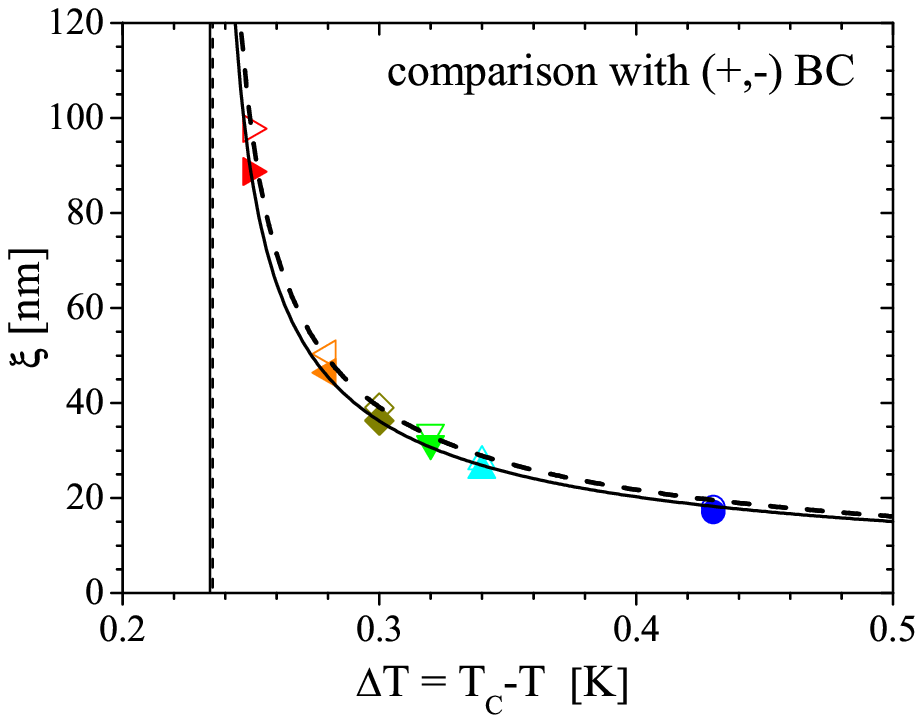}
\caption{%
Same as Fig.~\ref{fig:fitximm}, obtained by optimizing the agreement between
the experimental data from Fig.~\ref{fig:potpm} and the corresponding
theoretical predictions for the critical Casimir potential for the 
$\PM$ boundary
conditions (Fig.~\ref{fig:DerjPot}).  
}
\label{fig:fitxipm}
\end{figure}

For the $\PM$ boundary conditions, 
in Fig.~\ref{fig:potpm} we report the comparison between the theoretical
prediction and those experimental data which have been 
selected as explained above. As in
Fig.~\ref{fig:potmm}, the solid and dashed lines correspond to the predictions
based on Eq.~\reff{eq:PhiDgen} by using for $\Theta_{\PM} = \Theta_{\MP}$ 
the scaling functions indicated as solid or
dashed line, respectively, in Fig.~\ref{fig:DerjPot}. 
(As in Fig.~\ref{fig:potmm}, the experimental
potentials have been shifted vertically for each temperature 
$\Delta T_i$ by the amount $-\Phi_{{\rm offest},i}$ determined previously 
as the best fit parameter.)
The corresponding ensuing values $\xi_i$ of the
correlation length are reported in Fig.~\ref{fig:fitxipm}, 
together with the resulting best fit based on Eq.~\reff{eq:fitxi},
which leads to the least-square estimates 
$\xi_0^{\rm (fit)} = 1.8\pm 0.1$\AA\ and
$\Delta T_c^{\rm (fit)} \equiv T_c - T_c^{\rm (fit)} = 234\pm 2$mK 
for the solid curve and  
$\xi_0^{\rm (fit)} =  1.9\pm 0.1$\AA\ and
$\Delta T_c^{\rm (fit)} = 235\pm 2$mK
for the dashed one.
The final combined estimate of $\Delta T_c^{\rm (fit)}$, which takes into
account the systematic uncertainty of the amplitude of the 
theoretical prediction of $\Sf_{\PM}$, is
therefore $\Delta T_c^{\rm (fit)} = 235\pm 3$ mK. 
The correlation lengths reported in Fig.~\ref{fig:fitxipm} are in the range 
$20 \mbox{nm} \lesssim \xi \lesssim 95 \mbox{nm}$. The corresponding
ranges of distances $z$ depend on the temperature $\Delta T$ (see
Fig.~\ref{fig:potpm}) so that the experimental data cover the scaling variable
ranges $3.4 \lesssim x = z/\xi \lesssim 17$ and
$7 \lesssim u = (1-T/T^{\rm (fit)}_c)^{1/\nu}(z/\xi_0^{\rm
  (fit)})^{1/\nu} \lesssim 85$ (see Fig.~\ref{fig:DerjPot}).

\begin{figure}
\centering\includegraphics[width=\columnwidth]{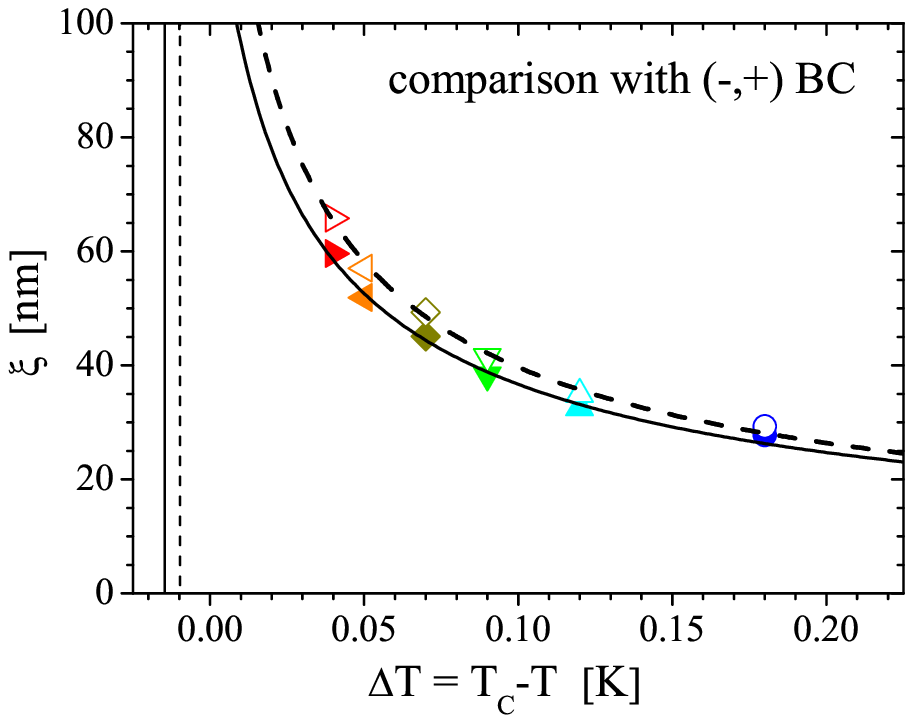}
\caption{%
Same as Figs.~\ref{fig:fitximm} and~\ref{fig:fitxipm}, 
obtained by optimizing the agreement between
the experimental data from Fig.~\ref{fig:potmp} and the corresponding
theoretical predictions for the critical Casimir potential for the 
$\MP$ boundary
conditions (Fig.~\ref{fig:DerjPot}).  
}
\label{fig:fitximp}
\end{figure}

For the $\MP$ boundary conditions, 
in Fig.~\ref{fig:potmp} we report the comparison between the theoretical
prediction and those experimental data which have been 
selected as explained above. As in
Figs.~\ref{fig:potmm} and~\ref{fig:potpm}, the sets of solid and dashed
lines, barely distinguishable, correspond to the predictions based on
Eq.~\reff{eq:PhiDgen} by using for $\Theta_{\MP} = \Theta_{\PM}$ 
the scaling functions indicated as solid or
dashed line, respectively, in Fig.~\ref{fig:DerjPot}.
(As in Figs.~\ref{fig:potmm} and~\ref{fig:potpm}, 
the experimental
potentials have been shifted vertically for each temperature 
$\Delta T_i$ by the amount $-\Phi_{{\rm offest},i}$ determined previously as
the best fit parameter.)
The corresponding values $\xi_i$ of the
correlation length, which can be inferred from this comparison, 
are reported in Fig.~\ref{fig:fitximp}, 
together with the resulting best fit based on Eq.~\reff{eq:fitxi},
which leads to the least-square estimates 
$\xi_0^{\rm (fit)} = 2.55\pm 0.25$\AA\ and
$\Delta T_c^{\rm (fit)} \equiv T_c - T_c^{\rm (fit)} = 14\pm 11\,$mK 
for the solid curve and 
$\xi_0^{\rm (fit)} = 2.7\pm 0.2$\AA\ and
$\Delta T_c^{\rm (fit)} \equiv T_c - T_c^{\rm (fit)} = 10\pm 9\,$mK 
for the dashed one. The final combined estimate of $\Delta T_c^{\rm (fit)}$,
which takes into account the systematic uncertainty of the amplitude of the
theoretical prediction of $\Sf_{\MP}$, is therefore $\Delta T_c^{\rm (fit)} =
 14\pm 11\,$mK. %
The correlation lengths reported in Fig.~\ref{fig:fitximp} are in the range 
$28\,\mbox{nm} \lesssim \xi \lesssim 66\, \mbox{nm}$ with the corresponding
ranges of the distances $z$ depending on the temperature $\Delta T$ (see
Fig.~\ref{fig:potmp}) so that the experimental data cover the scaling variable
ranges $7 \lesssim x = z/\xi \lesssim 67$ and
$22\lesssim u = (1-T/T^{\rm (fit)}_c)^{1/\nu}(z/\xi_0^{\rm (fit)})^{1/\nu}
\lesssim 790$ (see Fig.~\ref{fig:DerjPot}). 

The experimental data reported in Fig.~\ref{fig:potmp} (which
Fig.~\ref{fig:fitximp} refers to) have actually been acquired by
an experimental setup which makes use of 
an improved temperature control compared to
the one used during the acquisition of the data reported in
Figs.~\ref{fig:potmm}, \ref{fig:potpm}, \ref{fig:potpp}, 
and~\ref{fig:offcrit}. This upgrade of the setup is characterized by 
a better temperature stability and allows one to determine 
$T_c$ with higher accuracy~\cite{nhb-09}.
As far as the value of $\xi_0$ is concerned, taking into account the values
reported above for the $\MM$ and $\PM$ boundary conditions, one obtains the
combined estimate 
$\xi_0^{\rm (fit)} = 1.8 \pm 0.2$ \AA\  \cite{nature}
which is in very good agreement with the experimental value 
$\xi_0^{\rm (exp)}$ reported in the first line of 
Tab.~\ref{tab:xi0} (and quoted in Ref.~\cite{nature}).
It is interesting to note that the principal source of error in these estimates
of $\xi_0$ is actually the systematic uncertainty in the theoretical
predictions, which turns out to be more significant than the statistical or
possible systematic experimental errors, such as the one due to possible
variations or fluctuations of the temperature occurring during the measurement. 
Actually, due to the pronounced dependence of the theoretical predictions on
the temperature via the correlation length, these variations should result in
averaged effective values of the correlation length, most probably affecting
the overall amplitude, i.e., the value of $\xi_0^{\rm (fit)}$.
The estimate of $\xi_0^{\rm (fit)} = 2.6\pm 0.3\,$\AA\ 
based on the data for the
$\MP$ boundary conditions agrees with the estimate reported in the third line
of Tab.~\ref{tab:xi0} but it is larger and not 
quite compatible with the former one.
This might be due to larger systematic errors in the latter or due to possibly
different conditions of the mixture employed in the experiment (e.g., purity 
or possible contamination by leaching).
The combined estimate which accounts
for the results of our analysis is therefore
\be
\xi_0^{\rm (fit)} = 2.2\pm0.6\, \mbox{\AA},
\label{eq:xi0fit}
\ee
which is compatible and similar to the 
experimental value $\xi_0^{\rm (exp)}$ reported in Eq.~\reff{eq:xi0exp} but
carrying a larger, mainly systematic, uncertainty.
This agreement is particularly 
significant if one takes into account the fact that
Eq.~\reff{eq:xi0fit} combines results obtained from different experimental
conditions (different particles and different 
surface treatments), interpreted on the basis of the
available theoretical predictions. 
It is worthwhile to point out that one can equally reverse the line of
reasoning given above. One can adopt the point of view that the correlation
length is a \emph{bulk property} which is determined by independent bulk
measurement yielding Eqs.~\reff{eq:xidiv} and~\reff{eq:xi0exp}. This way $\xi$ 
is not a fit parameter but an input which fixes the theoretical prediction for
the critical Casimir potential completely. Since $\xi_0^{\rm (exp)}$ equals  
$\xi_0^{\rm (fit)}$, this implies that with this fixed input the theoretical
predictions yield the solid and dashed lines in Figs.~\ref{fig:potmm},
\ref{fig:potpm}, and~\ref{fig:potmp}. In this sense one can state that the
remarkable agreement between theory and experiment observed in the figures
occurs without adjusting parameters. 

\subsection{Noncritical composition} 
\label{sec:res:ncc}

\begin{figure}
\centering\includegraphics[width=\columnwidth]{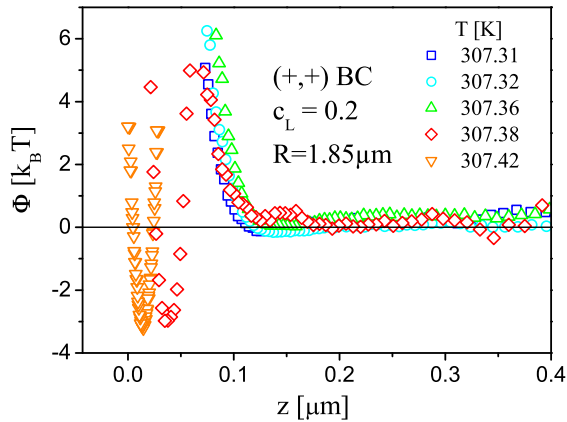}
\caption{%
Effective 
interaction potential $\Phi(z)$ between a wall and a particle of diameter 
$2R = 3.69\,\mu$m, immersed in a
water-lutidine mixture at mass fraction $c_L=0.2 < c_L^c$, as a function of
the distance $z$ and for various values of the temperature close to the
demixing phase boundary~\cite{nature}. This system corresponds to the 
$\PP$ boundary conditions. 
The gravitational and offset contributions to the potential
[see Eq.~\protect{\reff{eq:potgen}}] have been subtracted.
The abrupt formation of a narrow minimum of the potential close to the wall
upon increasing the temperature is interpreted as the formation of a liquid
bridge between the particle and the wall (see the main text).
}
\label{fig:offcrit}
\end{figure}

In the sense of renormalization-group theory the deviation of the bulk
concentration $c_L$ from its critical composition $c_L^c$ represent \emph{the}
second relevant scaling field besides the reduced temperature. In the language
of the Ising universality class the field conjugate to this deviation plays
the role of a bulk magnetic field. (Here we do not discuss that actually two
linear combinations of the conjugate field and of the reduced temperature form
the appropriate orthogonal scaling fields.)
In this sense, 
as already mentioned in Sec.~\ref{sec:th:nc},  
measurements along thermodynamic paths of varying temperature at fixed
off-critical compositions probe the dependence of the critical Casimir forces
on another, equally important scaling variable. Corresponding 
theoretical  predictions have been derived for the
parallel-plate geometry~\cite{f-th} as well as for two adjacent spherical
particles~\cite{colloids1a,colloids1b};
 one expects that a similar behavior holds 
for the present geometry of a sphere near 
a planar surface.
For a composition of the binary liquid mixture 
far away from its critical value
the critical Casimir forces become negligible.
For small deviations 
from the critical lutidine mass fraction $c_L^c$ and for  the
$\PP$ BC (so that both surfaces preferentially adsorb lutidine) 
 the temperature variation of the Casimir force upon approaching 
the two-phase coexistence line near the critical
point from the mixed phase  depends on whether 
$c_L$ is larger or smaller than   $c_L^c$. If $c_L$ is slightly less than
the critical composition, $c_L \lesssim c_L^c$, the Casimir force is expected to  behave
similarly as along the critical composition, i.e., 
it should exhibit a minimum before reaching the temperature 
$T_*(c_L)$ at which phase separation occurs [$T_*(c_L=c_L^c) = T_c$]. 
However,  the  depth of the critical Casimir potential is expected to  be 
considerably deeper away from the critical composition (see Fig.~8
in Ref.~\cite{colloids1b}). For lutidine mass fractions slightly
larger then the critical value, $c_L > c_L^c$,  as  function of temperature the Casimir force
is expected to vary similarly as in the
case  $c_L < c_L^c$  but 
much weaker.

These expectations are in agreement with the observations made 
in our experiment for  the $\PP$ BC and for several values
of  lutidine mass fractions in the range $0.26 < c_L < 0.32$ [not shown].
For $c_L$ close to $c^c_L$ the measured potentials between the wall and the colloidal particle
look similar to those  obtained for the critical composition.  For $c_L$ much smaller than $c^c_L$
the potentials are similar to the ones 
shown in Fig.~\ref{fig:offcrit} (see, e.g., Fig.~6.9 in Ref.~\cite{h-th},
corresponding to $c_L\simeq 0.25$).

For the values of $c_L$ further away from $c_L^c$ the system is no longer 
near criticality. Therefore the critical Casimir force ceases to influence
 effective interactions between the colloidal particle and the wall.
However, for the $\PP$ BC and for temperatures allowing for phase separation, 
i.e., in the present case above the critical temperature, 
one expects a bridging transition to occur at compositions $c_L < c_L^c$.
If the lutidine concentration is below its critical value $c_L^c$ the conjugate
\emph{bulk} field favors the water-rich phase whereas for the $\PP$ BC the
confining surfaces prefer the lutidine-rich phase. At a single wall this competition gives rise
to wetting phenomena  and  in confined geometries to  condensation phenomena.
As discussed in detail in Sec.~\ref{sec:th:nc}, 
if a bridge of the condensing
phase connects two adjacent 
spheres immersed in the binary liquid mixture, 
there is an attractive wetting-induced interaction that pulls
the sphere together. We expect a similar scenario to occur for
the present geometry, i.e.,  if a wetting
bridge of the phase which is favored by both surfaces is formed between
 the spherical particle and the planar wall.

Indeed, the measurements carried out 
for  lutidine mass fractions $c_L\lesssim 0.2$
indicate that such a bridge formation takes place.  
In Fig.~\ref{fig:offcrit}
we plot the measured particle-wall interaction potentials at several
temperatures near but below the temperature of demixing $T_*(c_L=0.2)$, which
could not be located with sufficient accuracy. Moreover, the temperature
scale in Fig.~\ref{fig:offcrit} has not been calibrated with a reference
temperature, so that only temperature changes are significant.
As one can see for temperatures between 307.31 and 307.36\,K,
the potentials are very well described by Eq.~\reff{eq:potgen} 
which accounts for the electrostatic and gravitational contributions only.
However, upon  further increasing the temperature by 20\,mK we observe a 
markedly different behavior of the potentials. Suddenly the interaction
potentials are shifted  towards smaller values of distances $z$. Also
the shape is changed in that the potentials exhibit a narrow and deep minimum. 
This sudden shift of the potential well towards the surface 
indicates the onset of an attractive interaction between the particle and 
the wall. A further slight increase in temperature 
gives rise to an even stronger shift of the potential minimum 
towards the wall.
This phenomenon 
is observed only for  lutidine mass fractions smaller than the
critical value, i.e., on that side of the phase diagram where the mixture is
poor in the component that is preferentially adsorbed by both surfaces. This
behavior is
in stark contrast to critical Casimir forces which vary gradually as function
of the thermodynamic variables. 
On the other hand, for the $\MM$ boundary conditions and $c_L \gtrsim c_L^c$ 
the resulting potentials do not differ qualitatively from those shown in 
Fig.~\ref{fig:potmm}. But  by further increasing the concentration of
lutidine to $c_L=0.4$ one observes the sudden formation of 
a narrow and deep potential well upon
increasing the temperature (see the potential corresponding to
$c_L=0.3$, $0.32$, and $0.4$, reported in Fig.~6.11 of Ref.~\cite{h-th}).
These observations are in agreement with the theoretical concepts described
in Sec.~\ref{sec:th:nc}. 
The effective potentials associated with the formation of a  bridge, formed by the phase coexisting with the bulk phase,
between the particle and the substrate are theoretically expected to 
exhibit hysteresis upon changing the temperature back towards its start
value. However, in the present experiment, only rather weak hints for
this hysteresis have been  observed and actually no convincing evidence for it could be
produced.

\section{Summary, conclusions, perspectives, and applications}  
\label{sec:conc}

\subsection{Summary}

We have presented a detailed account of the experimental and theoretical 
investigations of
the effective forces acting on spherical colloidal particles of radius $R$
close to a substrate and immersed in a near-critical
binary liquid mixture, shortly reported in
Ref.~\cite{nature}. Based on total internal reflection microscopy
(Figs.~\ref{fig:data_acq}, \ref{fig:data_analy}, and~\ref{fig:Dsenkrecht}) 
our main experimental findings are the following:
\begin{enumerate}
\item Upon raising the temperature $T$ of the binary liquid mixture of water
  and lutidine at its critical concentration towards its lower critical point
  $T_c$ of demixing (see Fig.~\ref{fig:WL-PD}), 
an {\it attractive} or {\it repulsive}
  force acting on the colloidal particle arises gradually. 
\item This effective force is attractive if the surfaces of the colloid
  and of the 
substrate display preferential adsorption of the same component of
  the mixture (see Figs.~\ref{fig:potmm} and~\ref{fig:potpp}), whereas it is
  repulsive in the cases of opposing preferences (see Figs.~\ref{fig:potpm}
  and~\ref{fig:potmp}). 
  This contribution to the total effective force (compare
  Sec.~\ref{sec:res:zp} and
  Fig.~\ref{fig:epsilons})
  is negligible at temperatures a few hundred mK away from $T_c$ and
  it increases significantly upon approaching it. 
  As experimentally verified, these so-called 
  critical Casimir forces
  can be reversibly switched on and off by changing
  the temperature.
\item If the concentration of 
  this binary liquid mixture is close to but not
  equal to the critical one we have observed a
  behavior which is qualitatively similar to the one observed for the mixture
  at its critical
  concentration. In the close 
  vicinity of the critical point there is no
  experimental evidence for the occurrence of  wetting phenomena. 
\item If the concentration of the binary liquid mixture differs significantly
  from the critical one and both surfaces exhibit the same 
  preferential adsorption for that
  component of the mixture which is disfavored in the bulk, 
  we observe the abrupt formation of a narrow and
  deep potential well (see Fig.~\ref{fig:offcrit})  upon approaching the
  phase boundary of first-order demixing. 
\end{enumerate} 
The experimental observations 1, 2, and 3 
can be consistently interpreted in terms of the
occurrence of the critical Casimir effect in near-critical mixtures, 
whereas
observation 4 can be understood in terms of 
the formation of a bridgelike configuration of the segregated phases 
(see Fig.~\ref{fig:phasediagram}).
For mixtures at the critical concentration it is
possible to quantitatively
compare the measured potentials with the corresponding
theoretical predictions for the contribution of critical Casimir forces 
(see Eqs.~\reff{eq:scalF} and \reff{eq:PhiDgen} as well as  
Figs.~\ref{fig:DerjPot} 
and~\ref{fig:DerjFor}), derived within the Derjaguin approximation (see
Fig.~\ref{fig:Derj}) and for the range of
distances within which electrostatic forces are negligible (see the solid
lines in Figs.~\ref{fig:potmm}, \ref{fig:potpm}, and~\ref{fig:potmp}). 
The correlation length $\xi$, 
as determined from the comparison between the
experimental data and the
theoretical predictions, follows rather well the theoretically expected
universal power-law behavior (see Figs.~\ref{fig:fitximm}, \ref{fig:fitxipm},
and~\ref{fig:fitximp}) and the associated non-universal amplitude $\xi_0$ 
is in
agreement with previous independent experimental determinations for this
specific binary mixture [compare Eq.~\reff{eq:xi0fit} to Eq.~\reff{eq:xi0exp}
  and see Tab.~\ref{tab:xi0}]. 

The {\it same} 
critical Casimir forces, investigated here by using a water-lutidine
mixture, are expected to act on a colloid immersed in \emph{any} binary
liquid mixture close to its demixing point (or in any fluid close to its
gas-liquid critical point) 
and in the vicinity of a substrate. 
While the values of non-universal parameters, such as $T_c$ and $\xi_0$,
depend on the specific mixture, the resulting critical Casimir force is
described by a material-independent, \emph{universal} scaling behavior (see
Eq.~\reff{eq:Ksp}) and scaling function (see Eq.~\reff{eq:KDerj} and
Fig.~\ref{fig:DerjFor} for small particle-substrate separation) which depends
only on whether the adsorption preferences of the particle and of 
the substrate are equal $[\PP,\MM]$ or opposite $[\PM,\MP]$. 

\subsection{Discussion}

The experimental observations summarized above 
might contribute to the understanding 
of the \emph{reversible} 
aggregation of a dilute suspension of colloidal particles immersed in
a water-lutidine mixture close to its demixing point, 
which has been the subject of several 
experimental studies since it was first observed
in 1985~\cite{beysens1} (for a review see Ref.~\cite{beysens2}). The
formation of
pre-wetting layers around the particles was first invoked as a possible
explanation for this phenomenon. Later on it was experimentally 
demonstrated that aggregation might actually occur in a region of the phase
diagram which extends too far from the two-phase coexistence line and from the
wetting transition to be possibly related to pre-wetting
phenomena~\cite{gkm-92}. (However, no aggregation was observed for mixtures at
the critical concentration~\cite{gkm-92}.) 
Among the possible different mechanisms
(see, e.g., Ref.~\cite{PLB-98} for a summary) which might contribute to
explain this flocculation, also critical Casimir forces have been invoked 
theoretically, as summarized and discussed in 
Refs.~\cite{colloids1a,colloids1b,colloids2}. 
In particular, the experimental
observation (besides for 2,6-lutidine and normal 
water as solvent reported also for
colloids dispersed in mixtures of 3-methylpyridine, heavy, and normal
water~\cite{nkg-93,saxs} or 2-butoxyethanol and normal water~\cite{gw-97})
that flocculation phenomena are enhanced near but off 
the critical point, at compositions which are slightly poorer
in the component preferentially adsorbed by the colloids than the
critical one, matches with the fact that also the critical Casimir forces
attain their maximum values there. 
Although flocculation involves the interaction of {\it many} 
colloidal particles and therefore is a 
many-body phenomenon (which can also be interpreted as a genuine phase
transition in a ternary mixture~\cite{jk-97}), 
some features such as
the experimentally determined 
asymmetry of the aggregation line with respect to the critical concentration 
(see, e.g.,
Refs.~\cite{beysens1,gkm-92}) can be qualitatively accounted for by the
behavior of the effective 
interaction among two colloids, mediated by the near-critical
solvent. This problem was theoretically investigated in
Refs.~\cite{colloids1a,colloids1b,colloids2} within various approximations.
In this context, our experimental study of the interaction
between a {\it single} colloidal particle and a substrate 
suggests that the actual
magnitude of the forces due to the critical Casimir effect and due to the
formation of a bridgelike configuration 
are large enough to play an important role also
in aggregation phenomena. This has been demonstrated recently on patterned 
substrates~\cite{patt-exp,Tr-09}. However, more quantitative corresponding 
statements require additional dedicated studies of many-body effects.
We note that, depending on the specific physical and chemical properties
of the colloidal suspension under consideration, 
various mechanisms might be at play in determining its aggregation,
especially for charged colloids, for which screening effects or even
field-induced phase separation of the mixture 
might be predominant~\cite{PLB-98,Tso-07}.
In this respect, recent experimental studies of the structure factor of such
an aggregating colloidal suspension via synchrotron small angle X-ray scattering
(SAXS)~\cite{saxs} might provide important insight into the physical
mechanisms at play in that phenomenon.

At the critical composition, 
the present experimental study detected the occurrence of critical Casimir
forces in a range of distances $z$ which corresponds to 
a scaling variable $x=z/\xi \gg 1$. In this limit, some of the qualitative
features of the scaling function of the force (such as the occurrence of a
maximum for $\PP$ boundary conditions --- see Fig.~\ref{fig:DerjFor}) have
not been probed. 
Actually, in this limit the associated potential for $(+,\pm)$ boundary
conditions is very well described by an exponential function
$\Phi_\Cas(z)/(\kB T) \simeq 2\pi A_\pm (R/\xi)\exp(-z/\xi)$ 
[see Eqs.~\reff{eq:PhiDgen} and~\reff{eq:sfasy}]. A clear signature of the
collective nature of such an
interaction is the fact that its range 
is set by the correlation length $\xi$. On
the other hand, the functional form of this dependence on $z$ is quite generic
and actually is common to interactions of rather different nature (e.g.,
electrostatic ones -- see Eq.\reff{eq:el-rep} -- 
the range of which is set by the
screening length $\kappa^{-1}$). For a relatively small correlation
length $\xi\simeq \kappa^{-1}$ the critical Casimir interaction and the
electrostatic repulsion have the same range and, depending on the specific
values of the parameters, one of them might dominate over the other. 
Especially in this case one expects an interplay between these two effects
due to the fact that the order parameter profile develops inhomogeneities on
a length scale $\xi$ which is comparable to the typical length
$\kappa^{-1}$ which characterizes the electrostatic screening in the
\emph{homogeneous} medium. Taking into account that due to the
segregation of the mixture the ions
have different solubilities in water and lutidine, the enhancement of
one of these two components close to the confining surfaces might result in a
change of the screening of the electrostatic interaction compared to the case
of a non-critical homogeneous medium. 
Analogously, as dispersion forces depend sensitively on the spatial structure
of the dielectric media forming the system and on the associated 
permittivities $\perm(\omega,{\bf x})$, the inhomogeneities which build
up in the medium upon increasing $\xi$ might affect significantly the
background van-der-Waals contribution 
$\Phi_{0,{\rm vdW}}$ to the total potential 
compared to the estimate we gave for a
homogeneous mixture [see Eqs.~\reff{eq:vdW} and~\reff{eq:Ham}]. 
In contrast to the critical Casimir effect, however, a quantitative analysis
of the interesting interplay between critical fluctuations and
dispersion/electrostatic forces necessarily requires the knowledge of several
system-specific properties such as the actual spatially varying 
composition of the mixture and the resulting permittivity 
$\perm(\omega,{\bf x})$.
Some of these properties might be inferred experimentally via,
e.g., surface plasmon spectroscopy of the binary mixture 
close to the substrate.   
The comparison of the experimental evidences presented here 
with the theoretical
predictions has not generated an actual need to account for the possible 
interplays mentioned above. Presumably they result into effects
which are negligible in the range of variables explored in our experiment and
within our experimental accuracy. 
This might not be the case for different choices of the
particle and the mixture for which, e.g., dispersion forces and therefore their
possible modification due to critical fluctuations might be more relevant than
in the system investigated here. 
%

In order to compare the experimental
data with the  theoretical predictions
we have inferred the bulk correlation length $\xi$ 
from the experimental data for the critical Casimir potential 
(see Sec.~\ref{sec:res:cc}).
Reversing the line of
argument, $\xi$ can be inferred on the basis 
of the deviation $\Delta T$ from the
critical temperature and of the knowledge of the non-universal amplitude
$\xi_0$ which has been determined by independent bulk experiments. 
However, one could also determine the \emph{scaling functions} of the critical
Casimir potentials on the sole basis of experimental data, without need of any
additional theoretical information. 
This requires the experimental determination of the actual 
correlation length $\xi$ corresponding to each temperature, which can be
accomplished,
e.g., by studying the light scattered by the mixture sufficiently far from the
colloid and the substrate, probing the behavior of bulk
fluctuations.
Such an independent experimental determination of the scaling function would
provide an additional, valuable test of the theoretical predictions.

\subsection{Perspectives and applications}

%
Suitable chemical treatments of the surfaces in contact with the binary
liquid mixture can be used in order to control 
the strength of their preferential adsorptions (i.e., the corresponding
surface fields, see Sec.~\ref{sec:th:cc}) and therefore the
resulting critical Casimir force. 
The experimental data presented here have been consistently interpreted in
terms of the theoretical predictions corresponding to strong preferential
adsorption. However, in view of possible applications of these effective 
forces, it is also important to study in detail both 
experimentally~\cite{patt-exp}
and theoretically~\cite{patt-chem,M-08,SD-08,Tr-09} cases in which such a
preference is weaker or spatially modulated in a controllable fashion via
suitable chemically patterned substrates.
In the latter case and depending on the symmetry of the pattern, the resulting
critical Casimir force acting on the colloidal particle acquires a
\emph{lateral} component in addition to the normal one investigated
here~\cite{patt-exp,Tr-09}. 
This lateral force, as the normal one, is characterized 
by a universal scaling behavior and its range is again set by the correlation
length $\xi$, such that 
it can be switched on and off by controlling the distance
from the critical point. 
In addition, the force turns out to be rather sensitive to 
details of the imprinted chemical structure, e.g., the striped
pattern considered in Refs.~\cite{patt-exp,Tr-09}. A 
proper theoretical analysis of the  critical Casimir potential
enables one to infer from the experimental data knowledge about such details
even if they could not be determined by independent means, such as 
atomic force microscopy~\cite{Tr-09}.

The lateral Casimir force might also find applications in colloid
rheology. Consider, e.g., a dilute suspension of $(+)$ colloids exposed 
to a suitably fabricated
substrate which has an adsorption preference smoothly varying along one
direction from $(-)$ to $(+)$, such that it changes appreciably on the scale
of the radius of the colloid. For sufficiently small values of the
correlation length, the colloids diffuse isotropically along the substrate. 
However, upon approaching the critical point, the lateral Casimir force
associated with the spatial gradient of the preferential adsorption (i.e., of
the corresponding surface field) adds a deterministic drift to this diffusion
process, resulting in a transport of colloids along the surface of the
substrate. The direction of the flow will be reversed by changing the
preferential adsorption of the colloid from $(+)$ to $(-)$, which can
be exploited as a reversible selection mechanism. In this context, the
critical Casimir force acting on a micrometer-sized colloid exposed to a
substrate with a modulated adsorption preference on the 
the scale of some hundred micrometers has been recently studied
experimentally~\cite{nhb-09}.

Topographical modulations of the surface of an otherwise chemically
homogeneous substrate can also be used to control the direction
of the total force acting on a similar substrate~\cite{patt-top} 
or on a colloidal particle exposed to it. 
Additionally, chemical patterning or geometrical
deviations from spherical symmetry of the colloidal particle (e.g.,
ellipsoidal colloids) result in a  critical Casimir
\emph{torque}~\cite{eisen,khd-08} if the particle is exposed to a
substrate. Combining all these features one should be able, e.g., to control
reversibly via minute temperature changes the orientation of such colloids
exposed to a striped substrate. 
%

The critical Casimir force acting on a colloidal particle close to a plate
fluctuates in time due to the fact that it originates from
time-dependent critical fluctuations. In the present analysis we focused on
the \emph{mean} value of such a force and on the associated averaged
potential $\Phi_\Cas$. However, as explained in Sec.~\reff{sec:exp_TIRM}, 
TIRM naturally provides a measurement of the time-dependent sphere-plate
distance $z(t)$ which, in turn, can be used to determine the correlation
time $t_{\rm R}$ of the critical Casimir force and how its expected algebraic
temporal singularity builds up upon approaching the critical point. 
(Some aspects of this
dynamical behavior are discussed in Refs.~\cite{gd-06,g-08,gam-08}.)
This critical 
slowing down of the critical Casimir effect can in principle be exploited
in order to control the resulting dynamics of the colloidal particle.


%
In contrast to interactions which typically act among colloids (such as
electrostatic and dispersion forces), 
the critical Casimir force is characterized by 
a pronounced temperature dependence. This fact can possibly be exploited in
order to control via minute temperature changes the phase behavior and
aggregation phenomena in colloidal dispersions in the bulk or close to
those chemically structured solid surfaces which find applications in the
fabrication of nano- and micrometer scale devices. 
Not only the range of the resulting interaction can be controlled but also its
sign and spatial direction. This can be typically achieved by surface
treatments and it does not require (as it does, e.g., for dispersion forces) 
substantial changes or a fine tuning of the bulk 
properties of the materials
which constitute the immersed objects and the mixture itself.
These properties 
could be used, e.g., in order to neutralize the attractive quantum
mechanical Casimir force responsible for the stiction which brings
micro-electromechanical systems to a standstill. If these machines would work
not in a vacuum but in a liquid mixture close to the critical point,
the stiction could be prevented by tuning the critical Casimir force to be
repulsive via a suitable coating of the various machine parts. 
With optically removable or controllable coatings, one could
very conveniently control the functioning of the microdevice without acting
directly on it.
In addition,
properly designed surfaces might provide temperature-controlled confining
potentials which might find applications in self-assembly
processes~\cite{patt-exp}.

\begin{acknowledgments}
AG is grateful to Adrian Parsegian for useful
discussions on dispersion forces. 
AG, AM, and SD acknowledge the hospitality and support of the
Kavli Institute for Theoretical Physics at the University of Santa Barbara, 
within the program ``The theory and practice of fluctuation-induced
interactions'' under the grant No. NSF PHY05-51164 of the US National Science
Foundation. 
AG is supported by a MIUR fellowship within the program 
``Incentivazione alla mobilit\`a di studiosi stranieri e italiani residenti
all'estero''.
\end{acknowledgments}



\begin{thebibliography}{}
\bibitem{footnote_nvsI}
During the integration time $\Delta t$ the scattered intensity $I_{\rm sc}$
fluctuates due to the vertical diffusion of the colloid. Taking into
account Eq.~\reff{eq:TIRM} %
one has $n_{\rm sc}(t)/[I_{\rm sc}(t)\Delta t]
= (\Delta t)^{-1}\int_0^{\Delta t}\!\!\rmd\tau\exp\{-\zeta\Delta z(\tau;t)\}$
where $\Delta z(\tau;t) = z(t+\tau) - z(t)$ fluctuates as a function of $\tau$
for a fixed $z(t)$. Assuming that these fluctuations $\Delta z$ 
are small compared to $\zeta^{-1}$ one can expand the r.h.s.
and calculate its statistical average, taking into account that for short
times the motion of the particle is not significantly affected by the external
forces, so that $\langle [\Delta z(\tau;t)]^2\rangle \simeq 2 D_\bot \tau$
(independent of $t$),
where $D_\bot$ is the vertical diffusion coefficient of the colloid in the
mixture (c.f., Sec.~\ref{sec:exp_TIRM_da}). The result can be expressed in
terms of the root mean square displacement $\Delta z_{\rm rms} \equiv \langle
  [\Delta z(\Delta t;t)]^2\rangle^{1/2}$ of the colloid 
within the time interval $\Delta t$
  as $n_{\rm sc}(t)/[I_{\rm sc}(t)\Delta t] \simeq 1 + (\zeta \Delta z_{\rm
    rms})^2/4$. In order to provide an upper bound to this deviation from 1 
we assume
  $D_\bot\simeq D_\infty = \kB T/(6\pi\eta R) \simeq 0.09\,
  (\mu\mbox{m})^2/\mbox{s}$ for the colloid of smaller 
radius $R=1.2\,\mu$m in the
  near-critical mixture (c.f., Sec.~\ref{sec:exp_TIRM_da}). This results in 
$\Delta z_{\rm rms} \simeq 13\,\mbox{nm}$, i.e., $(\zeta\Delta z_{\rm
rms})^2/4 \simeq 10^{-3}$ so that $n_{\rm sc}(t)\simeq I_{\rm sc}(t)\Delta
  t$. Note that, for a given time interval $\Delta t$, 
$\Delta z_{\rm rms}$ provides also a
  lower bound to the accuracy with which one can determine the actual position
  of the colloid via TIRM.
\bibitem{footnote1}
The statistical uncertainty $\Delta z$ of the value of $z$ due to the
counting statistics of the single photon counter can be estimated as $(\Delta
z)_{\rm stat} = \zeta^{-1} \Delta n_{sc}/n_{\rm sc}$ where, 
assuming a Poissonian
counting statistics, one has  
$\Delta n_{\rm sc} = n_{\rm sc}^{1/2}$. For the typical
measured number of photons $n_{\rm sc} \simeq 2\times 10^3$, 
with $\zeta^{-1} = 200\,$nm this leads to an uncertainty of $\Delta z \simeq
5\,$nm. 
This is negligible on the typical length scale $z\simeq 100$nm
we are interested in.
\bibitem{footnote_det_exp}
The surface treatments have been done either by exposing the cell to HMDS
vapor overnight or by rinsing it with 0.1M NaOH for 30 min.
\end{thebibliography}
\end{document}